\documentclass[a4paper,fleqn,usenatbib]{mnras}

\usepackage{newtxtext,newtxmath}
\usepackage[T1]{fontenc}
\usepackage{ae,aecompl}

\usepackage{graphicx}	
\usepackage{amsmath}	
\usepackage{bm}
\usepackage{cancel}
\usepackage{lipsum}

\usepackage{times}
\usepackage{threeparttable}
\usepackage{multicol} 
\usepackage{ulem}
\usepackage{booktabs}
\usepackage{comment}
\usepackage{multirow}
\usepackage{hyperref} 
\usepackage{pdfcolfoot}
\usepackage[title]{appendix}

\usepackage[dvipsnames]{xcolor}

\def\dmunit{{\rm pc\ cm^{-3}}}
\def\beq{\begin{equation}}
\def\eeq{\end{equation}}
\def\beqn{\begin{eqnarray}}
\def\eeqn{\end{eqnarray}}

\def\beq{\begin{equation}}
\def\eeq{\end{equation}}
\def\beqn{\begin{eqnarray}}
\def\eeqn{\end{eqnarray}}

\title[Outflows from the Galactic FRB]{Multi-Wavelength Constraints on the Outflow Properties of the Extremely Bright Millisecond Radio Bursts from the Galactic Magnetar SGR 1935$+$2154}

\author[Yamasaki, Kashiyama \& Murase]
{Shotaro Yamasaki\thanks{E-mail: shotaro.yamasaki@mail.huji.ac.il}$^{1}$, Kazumi Kashiyama$^{2,3}$, and Kohta Murase$^{4,5,6,7}$
\\
$^{1}$Racah Institute of Physics, The Hebrew University of Jerusalem, Jerusalem 91904, Israel\\
$^{2}$ Department of Physics, School of Science, The University of Tokyo, 7-3-1 Hongo, Bunkyo-ku, Tokyo 113-0033, Japan\\
$^{3}$ Research Center for the Early Universe, School of Science, The University of Tokyo, 7-3-1 Hongo, Bunkyo-ku, Tokyo 113-0033, Japan\\
$^{4}$ Department of Physics, The Pennsylvania State University, University Park, Pennsylvania 16802, USA
\\
$^{5}$ Department of Astronomy \& Astrophysics, The Pennsylvania State University, University Park, Pennsylvania 16802, USA
\\
$^{6}$
Center for Multimessenger Astrophysics, Institute for Gravitation and the Cosmos, The Pennsylvania State University, University Park, Pennsylvania 16802, USA
\\
$^{7}$ Center for Gravitational Physics, Yukawa Institute for Theoretical Physics, Kyoto, Kyoto 606-8502 Japan
}

\date{Accepted XXX. Received YYY; in original form ZZZ}

\pubyear{2020}

\begin{document}
\label{firstpage}
\pagerange{\pageref{firstpage}--\pageref{lastpage}}
\maketitle

\begin{abstract}
Extremely bright coherent radio bursts with millisecond duration, reminiscent of cosmological fast radio bursts (FRBs), were co-detected with anomalously-hard X-ray bursts from a Galactic magnetar SGR 1935$+$2154. We investigate the possibility that the event was triggered by the magnetic energy injection inside the magnetosphere, thereby producing magnetically-trapped fireball (FB) and relativistic outflows simultaneously. The thermal component of the X-ray burst is consistent with a trapped FB with an average temperature of $\sim200$--$300$ keV and size of $\sim10^5$ cm. Meanwhile, the non-thermal component of the X-ray burst and the coherent radio burst may arise from relativistic outflows. We calculate the dynamical evolution of the outflow, launched with an energy budget of $10^{39}$-$10^{40}$ erg comparable to that for the trapped FB, for different initial baryon load $\eta$ and magnetization $\sigma_0$. If hard X-ray and radio bursts are both produced by the energy dissipation of the outflow, the outflow properties are constrained by combining the conditions for photon escape and the intrinsic timing offset $\lesssim10$ ms among radio and X-ray burst spikes. We show that the hard X-ray burst must be generated at $r_{\rm X}\gtrsim10^{8}$ cm from the magnetar, irrespective of the emission mechanism. Moreover, we find that the outflow quickly accelerates up to a Lorentz factor of $10^2\lesssim\Gamma\lesssim10^3$ by the time it reaches the edge of the magnetosphere and the dissipation occurs at $10^{12}$ cm $\lesssim r_{\rm radio,X}\lesssim10^{14}$ cm. Our results imply either extremely-clean ($\eta\gtrsim10^4$) or highly-magnetized ($\sigma_0\gtrsim10^3$) outflows, which might be consistent with the rarity of the phenomenon.

\end{abstract}



\section{Introduction}
Recently, one of the most prolific transient magnetars, SGR J1935$+$2154 \citep{israel16} went into an intense bursting episode on April 27 2020, and hundreds of X-ray bursts were recorded in a few hours \citep{borghese20,nicergbm_paper}. 
During this active phase, an extremely intense radio burst with millisecond duration, reminiscent of cosmological fast radio bursts (FRBs), was detected by radio telescopes CHIME/FRB \citep{chime_paper} and STARE2 \citep{stare2_paper} on April 28 2020, strengthening the connection between FRBs and magnetars. 
Importantly, Insight/HXMT~\citep{hxmt_paper}, Konus-Wind~\citep{konus_paper}, INTEGRAL/IBIS~\citep{integral_paper} and AGILE~\citep{agile_paper} independently detected an X-ray burst associated with the FRB-like radio burst \citep{integral_paper,hxmt_paper,konus_paper,agile_paper}; 
the timing of the emissions is the same within the observational uncertainties and the radio burst detected by CHIME have a similar temporal structure to the X-ray burst. 
The X-ray burst is peculiar in that the spectrum is much harder than a typical SGR burst with comparable (or even higher) fluences and FRBs were not detected with many other X-ray bursts from the same source~\citep{fast_upperlimit_paper}.

Theoretical interpretations of the April 28 event are broadly classified into two categories: ``close-in" and ``far-away" scenarios, depending on how close the radio emission is generated from the central engine (i.e., the magnetar). The former includes the curvature radiation in the open magnetic fields (e.g., \citealt{lu20,katz20,yang21}), the plasma instability triggered by magnetic reconnection \citep{lyutikov20,lyutikov20b} and the low-altitude magnetospheric emission \citep{wadiasingh19,wadiasingh20}, whereas the latter invokes a maser-type instability at the shock between magnetar flare wind and the pre-existing material (e.g., \citealt{margalit20,yuan20,yu20}).
The possibilities of generating double/multiple-peaked radio pulses by the quasi-periodic oscillation of magnetars \citep{wang20} or the scintillation effect \citep{simard20} are also discussed. Whether the April 28 event as well as cosmological FRBs are generated by close-in or far-away models is subject to intense debate in the community, both from observational and theoretical aspects. Regarding cosmological FRBs, recent FAST observations on varying polarization angles in some repeaters \citep{luo20} and discovery of recurrent bursts from FRB 121102 with too short separations down to milliseconds and too large energy budget \citep{li21} to accommodate the far-away (maser-type)  models may prefer close-in (curvature-type) models. Meanwhile, it has also been pointed out that close-in models could have some theoretical flaws because realistic plasma effects are often neglected (e.g., \citealt{lyubarski21}).

In either the close-in and far-away models, the event is triggered by a deposition of magnetic energy in the magnetosphere, which may results in the formation of an electron/positron ($e^{\pm}$) plasma bubble confined to the stellar surface by the strong magnetic pressure \citep{td95,td96}, so-called trapped fireball (FB), and also launching an outflow of relativistic plasma (or an expanding FB). In this paper, we aim to put general constraints on such FBs (i.e., the properties of the outflow responsible for the X-ray and/or radio bursts) with modest assumptions on the radiation mechanism, based on the multi-wavelength observations of the April 28 event. 
While the thermal component of the X-ray burst is consistent with a trapped FB, the non-thermal component of the X-ray burst and the coherent radio burst may arise from relativistic outflows. Regarding the origins of hard X-ray burst, we examine the two possibilities that (1) it is produced in the vicinity of the NS or the trapped FB (\S \ref{ss:trappedFB}) and that (2) it arises from the relativistic outflow (\S \ref{ss:hard X-ray from outflows}). The second possibility is investigated by considering the evolution of outflows with different properties, assuming that the hard X-ray and coherent radio bursts have been produced due to some sort of energy dissipation inside the outflows, which broadly includes far-away models.

This paper is organized as follows. In \S\ref{s:FRB200428}, we summarize the key observational properties of the April 28 event. We constrain the total energy budget of the event in \S\ref{ss:trappedFB} by assuming that the thermal component of the X-ray spectrum is due to the trapped FB. In \S\ref{ss:outflow models}, we calculate the dynamical evolution of the outflow, which is likely responsible for the FRB-like burst and the non-thermal part of the hard X-ray burst spectrum. Constraints on the outflow properties are set from the general conditions required to generate the emission in \S\ref{s:outflow constraints} and our findings are summarized and implications are discussed in \S\ref{s:discussion}.
Hereafter, we use $Q_x\equiv Q/10^x$ in cgs units.

\section{Key Observed Properties of April 28 Events from SGR 1935+2154}
\label{s:FRB200428}
Here we review the key observed properties of the radio and X-ray bursts from SGR 1935$+$2154 on May 28 2020 (see also Table \ref{tab:SGR1935+2154}).

{\it SGR 1935+2154.--} 
SGR 1935$+$2154 is one of the most prolific transient magnetars;
the spin period and the spindown rate are measured to be $P_{\rm spin}=3.24$ s and $\dot P = 1.43\times 10^{-11}\,\mathrm{s\,s^{-1}}$, respectively~\citep{israel16}. Accordingly, the surface dipole magnetic field strength is estimated as $B_\mathrm{p}=2.2\times10^{14}$ G. 
This magnetar has been recently in an active phase since April 27 2020~\citep{nicergbm_paper}.
The distance estimate is somewhat uncertain.
SGR 1935$+$2154 is spatially associated with the supernova remnant (SNR) G57.2$+$0.8. Throughout this work, we adopt a source distance of $10$ kpc, which is consistent with the different distance estimates between $6.7$ kpc \citep{zhou20} and $12.5$  kpc \citep{kothes18} in the literature.

{\it Radio Observations.--}
The radio burst from SGR 1935$+$2154 was detected independently by CHIME at $400$--$800$ MHz and STARE2 at $1.4$ GHz~\citep{chime_paper,stare2_paper}. The CHIME burst consists of two sub-bursts with widths of $\sim5$ ms separated by $\sim30$ ms, whereas the STARE2 burst has a single narrow spike with a width of $0.61$ ms. 
According to the total fluence reported by STARE2, the radiated energy (isotropic equivalent) is estimated to be $E_{\rm radio}^{\rm iso}=(0.3$--$2.4)\times10^{35}\ (d/10\ {\rm kpc})^2$ erg. 
The observed dispersion measures (DM) in both radio observations are consistent with a single value, DM $\sim332.7\ \dmunit$\citep{chime_paper,stare2_paper}, which is in agreement with sources in the Galactic plane. 
Except for the detection of other low-luminosity radio events\footnote{Most recently, a pair of four-orders-of-magnitude less bright ($112\ {\rm Jy~ms}$ and $22\ {\rm Jy~ms}$) radio bursts with temporal separation of $1.4$ s at $1.32$ GHz has been discovered by a coordinated multi-telescope observation \citep{kirsten20}, albeit without X-ray (or gamma-ray) counterparts.}, the FAST set stringent upper limits on the radio flux associated with many other X-ray bursts~\citep{fast_upperlimit_paper}. 

{\it X-ray Observations.--} 
There are four co-detections of the hard X-ray burst associated with the FRB-like radio burst \citep{hxmt_paper,konus_paper,integral_paper,agile_paper}. 
The total duration of the burst is roughly $0.3$--$0.5$ s. The X-ray light curves consist of a few narrow peaks with each sub-burst width $\lesssim 10$ ms \citep{hxmt_paper,konus_paper,integral_paper}, which is coincident with the radio-burst arrival times (see below). The X-ray spectrum extends up to $250$ keV \citep{konus_paper,hxmt_paper} and is fitted by an exponentially-cutoff power law function with a typical peak energy $\epsilon_{\rm p}\sim50$--$100$ keV. This is unusually hard compared to other X-ray bursts with comparable (or even higher) fluence detected in the same \citep{nicergbm_paper} and past \citep{konus_paper,hxmt_paper,integral_paper} bursting episodes. There is evidence for a temporal spectral
hardening associated with two peaks of the burst  \citep{integral_paper,hxmt_paper}.
The isotropic energy in the X-ray bands is $E_{\rm X}^{\rm iso}=(0.5$-$1.2)\times10^{40}\ (d/10\ {\rm kpc})^2$ erg, which is $\sim 10^5$ times larger than the radio bands.

\begin{figure}
\centering
\includegraphics[width=.48\textwidth]{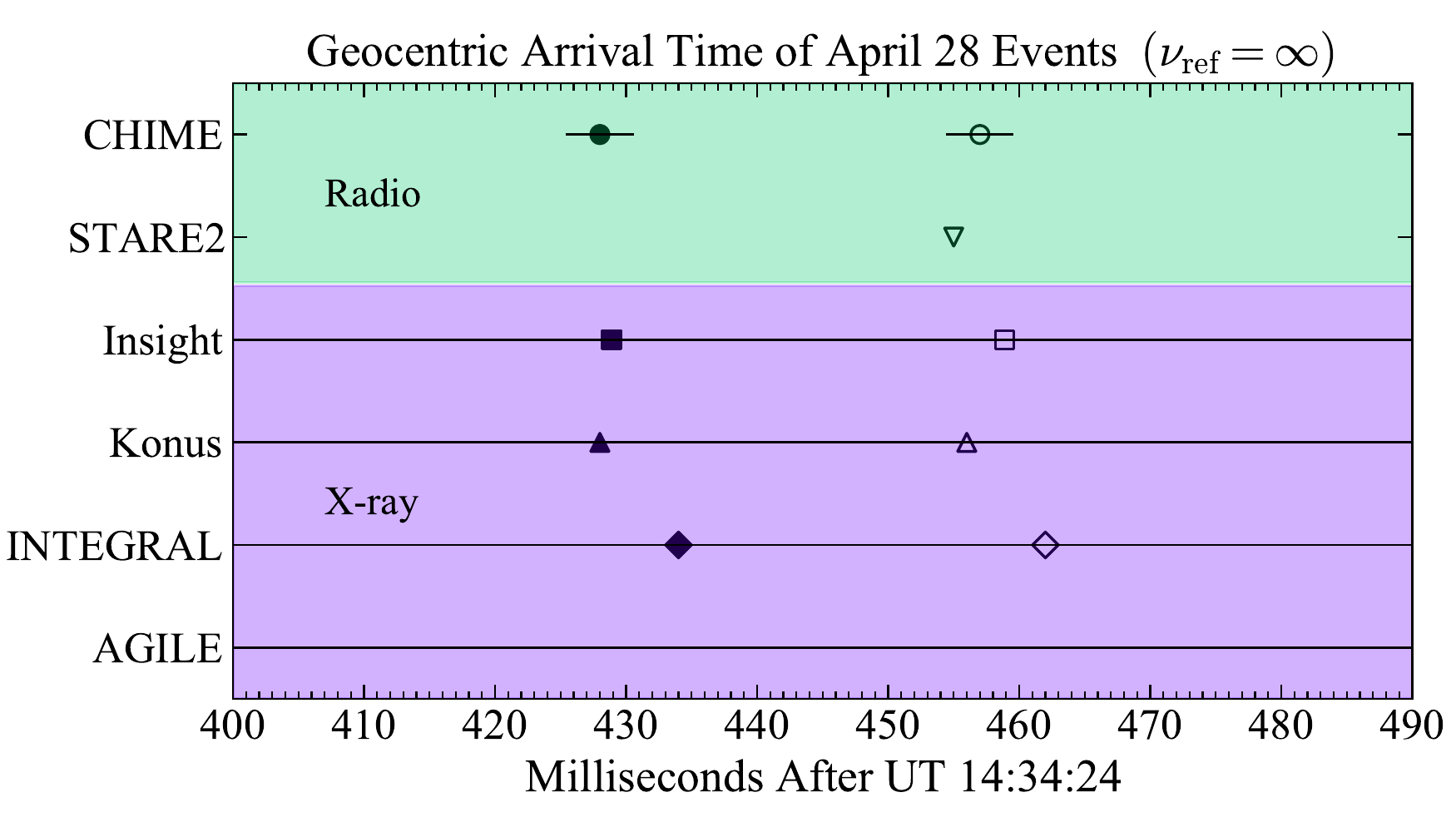}
\caption{Timelines of the radio and X-ray burst from SGR 1935+2154 on May 28 2020. The arrival time delay due to radio dispersion is subtracted assuming DM  $=332.7\,\mathrm{pc\,cm^{-3}}$ and $\nu_{\rm ref} = \infty$ (see Eq. [\ref{eq:t_DM}]). Each horizontal black bar represents the duration of individual burst. 
Peak information for AGILE is not available due to the relatively low temporal resolution of $\sim0.5$ s, and not shown here. }
\label{fig:TOA}
\end{figure}

{\it Burst Arrival Time.--}
The arrival time delay of a pulse with an observed frequency of $\nu$ with respect to a reference frequency $\nu_\mathrm{ref}$ is 
\beqn
\label{eq:t_DM}
t_{\rm DM}(\nu,\nu_\mathrm{ref})=k_{\rm DM}\left(\frac{1}{\nu^{2}}-\frac{1}{\nu_\mathrm{ref}^{2}}\right)\,{\rm DM},
\eeqn
where $k_{\rm DM}\equiv e^2/(2\pi m_e c)\simeq4.15\ {\rm ms\ pc^{-1}\ cm^3\ GHz^2}$ is the dispersion constant. 
The dispersion delay between CHIME and STARE2 is $t_{\rm DM}(600{\,\rm MHz},1.53{\,\rm GHz})\simeq3.25$ s, which is consistent with the observed time delay between the second CHIME sub-burst and the STARE2 burst (see Figure \ref{fig:TOA}). In fact, the spectrum of the second CHIME sub-burst extends up to higher frequency ($\sim800$ MHz), whereas the first CHIME sub-burst has an apparent spectral cutoff at $\lesssim600$ MHz \citep{chime_paper}. Furthermore, the spiky temporal structure of the second CHIME sub-burst resembles that of STARE2 burst. These all implies that the STARE2 burst may be of the same origin as the second CHIME sub-burst.

On the other hand, the dispersion delay between CHIME and the X-ray satellites is $t_{\rm DM}(600{\,\rm MHz},\infty)\sim 3.84$ s.
Given this, the arrival times of first/second CHIME sub-bursts and the first/second peaks in the X-ray light curves\footnote{After the refined analysis, the Integral light curve shows three narrow peaks \citep{integral_paper}. The third peak separated from the second one by $\sim31$ ms is not shown in Figure \ref{fig:TOA}.} are consistent within error of $\Delta t_{\rm CHIME,X} \equiv t_{\rm X}-t_{\rm CHIME}\lesssim5$ ms. Even if we additionally take into account the finite time-resolution of X-ray detectors ($\lesssim2$ ms around burst peaks) and pulse width of CHIME sub-bursts ($\sim5$ ms), most conservatively we get $\Delta t_{\rm CHIME,X}\equiv t_{\rm X}-t_{\rm CHIME}\lesssim10$ ms. Similarly, we obtain $\Delta t_{\rm STARE2,X}\equiv t_{\rm X}-t_{\rm STARE2}\lesssim10$ ms for the STARE2 burst and the X-ray second peak. 
In summary, the intrinsic time separation between X-ray and radio emission peaks is estimated to be no longer than $|\Delta t_{\rm X,radio}|\sim10$ ms. 

\begin{table*}
\caption{Properties of the radio and hard X-ray burst associated with Galactic magnetar SGR 1935$+$2154 on 2020 April 28.}
\begin{threeparttable}
\label{tab:SGR1935+2154}
\centering
\begin{tabular}{llllcccc}
\hline\hline
Band & Telescope & Frequency & Arrival Time UT ($\nu_{\rm ref}$)$^{\rm a}$& Total Duration & Total Fluence &Ref.$^{\rm b}$& Energy$^{\rm c}$\\\hline
\multirow{4}{*}{Radio}&\multirow{2}{*}{CHIME}&\multirow{2}{*}{$0.4$--$0.8$ GHz}&14:34:28.264 ($0.6$ GHz) &\multirow{2}{*}{40 ms$^{\rm d}$}  & \multirow{2}{*}{$700^{+700}_{-350}$ kJy ms }&\multirow{2}{*}{[1]}&\multirow{2}{*}{$3\times10^{34}$ erg}\\
&&&{14:34:24.428} ($\infty$) && &&\\
&\multirow{2}{*}{STARE2}&\multirow{2}{*}{$1.28$--$1.53$ GHz}& 14:34:25.046 ($1.53$ GHz) &\multirow{2}{*}{$0.61$ ms$^{\rm e}$}  & \multirow{2}{*}{$1.5^{+0.3}_{-0.3}$ MJy ms }&\multirow{2}{*}{[2]}&\multirow{2}{*}{$2.4\times10^{35}$ erg}\\
&&&{14:34:24.455} ($\infty$) && &&\\
\hline
\multirow{4}{*}{X/$\gamma$}& Insight-HXMT&$1$--$250$ keV& {14:34:24.429(2)} ($\infty$)$^{\rm f}$ &$\sim0.5$ s   & $7.1_{-0.4}^{+0.4}\times10^{-7}\ {\rm erg\ cm^{-2}}$ &[3]&$6\times10^{39}$ erg\\
&Konus-{\it Wind}&$20$--$500$ keV& {14:34:24.428(1)}  ($\infty$)$^{\rm e}$ & $\sim0.3$ s  & $9.7_{-1.1}^{+1.1}\times10^{-7}\ {\rm erg\ cm^{-2}}$ &[4]&$1.2\times10^{40}$ erg\\
&INTEGRAL&$20$--$200$ keV& {14:34:24.434}  ($\infty$)$^{\rm e}$ &$\sim0.3$ s & $5.2_{-0.4}^{+0.4}\times10^{-7} {\rm erg\ cm^{-2}}$ &[5]&$5\times10^{39}$ erg\\
&AGILE&$18$--$60$ keV& 14:34:24.4 ($\infty$) &$\lesssim0.5$ s & $5\times10^{-7}\ {\rm erg\ cm^{-2}}$ &[6]&$5.6\times10^{39}$ erg\\
\hline\hline
\end{tabular}
 \begin{tablenotes}[para,flushleft,online,normal] 
 \footnotesize
  \small
  \item[a]{Geocentric arrival time of the first peak at reference frequency $\nu=\nu_{\rm ref}$ with ${\rm DM}=332.7\ \dmunit$;}
  \item[b]{[1] \citet{chime_paper} [2] \citet{stare2_paper} [3] \citet{hxmt_paper} [4] \citet{konus_paper} [5] \citet{integral_paper} [6] \citet{agile_paper};}
  \item[c]{Assuming a distance of $10$ kpc;}
  \item[d]{The event consists of two sub-bursts with widths of $\sim5$ ms separated by $\sim30$ ms;}
  \item[e]{A single spiky burst;}
  \item[f]{Bursts have complicated temporal structure with multiple narrow peaks and here the geocentric arrival time of the first peak is shown}
  \end{tablenotes}
  \end{threeparttable}
\end{table*}

\section{Theoretical Framework}
Whatever the emission mechanism of the the radio burst is, (i) the event is likely to be triggered by an injection of energy into the magnetosphere, and (ii) the radio emission may arise from a relativistic outflow at a sufficiently large distance from the NS surface in order to avoid a significant scattering and absorption.
The launch of a relativistic outflow 
might be accompanied by the formation of a trapped FB \citep{td95}. 
We first constrain the energy and size of the trapped FB from the X-ray data, and then calculate the expansion of the outflow, assuming that the energy and size of the outflow at launch are comparable to those of the trapped FB.

\subsection{Trapped Fireball}
\label{ss:trappedFB}
\begin{figure}
\centering
\includegraphics[width=.48\textwidth]{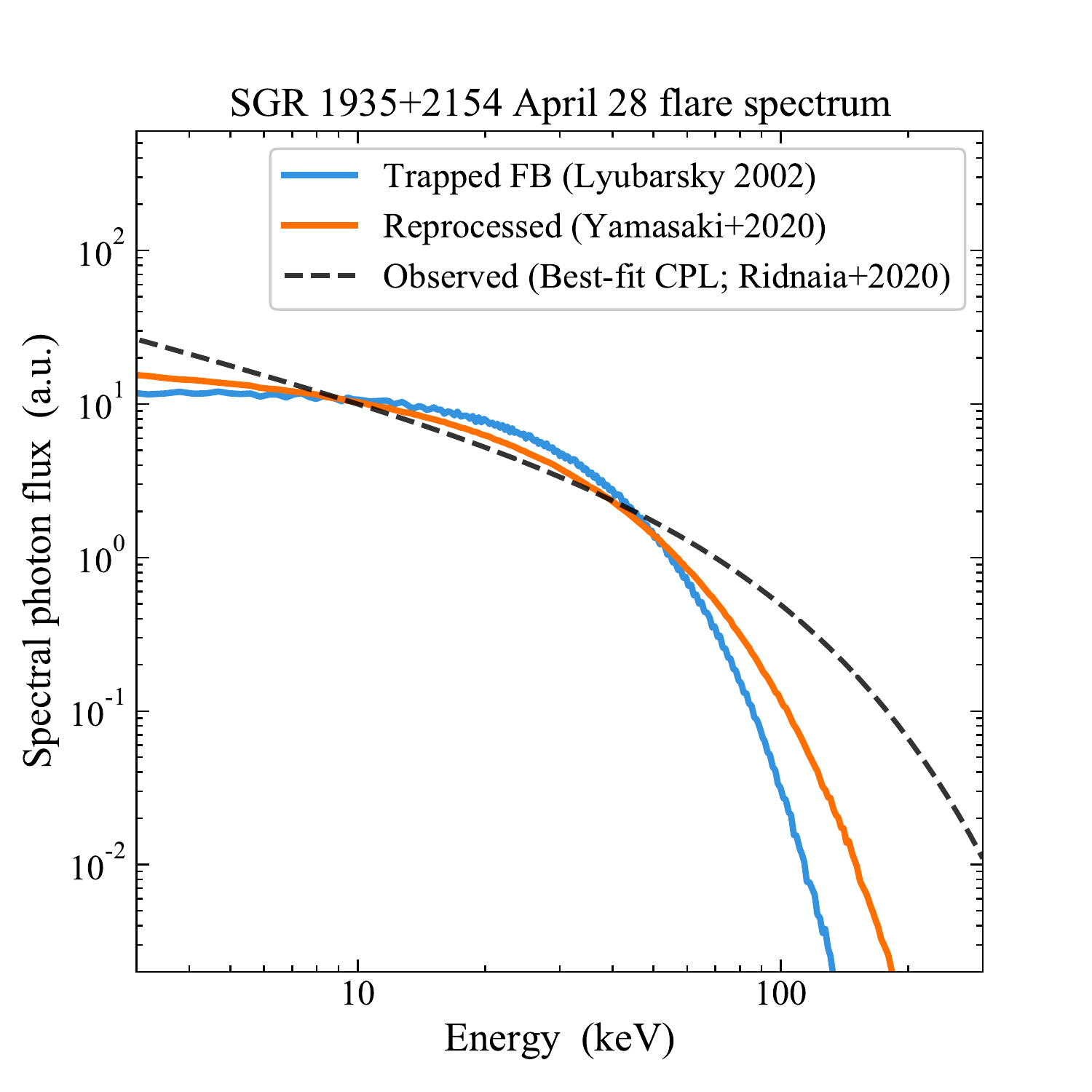}
\caption{Resonant cyclotron scattering spectra that might be sampled during magnetar flares (\citealt{yamasaki20}, orange solid line). The seed photon spectrum (the modified blackbody spectrum proposed by \citealt{lyubarsky02}) with an effective temperature of $10$ keV is also shown by the blue solid line. The best-fit exponentially-cutoff power law (CPL) function ($dN/d\epsilon\propto\epsilon^{\alpha}\exp{\left[-(\alpha+2)(\epsilon/\epsilon_{\rm p})\right]}$ with $\alpha=-0.72_{-0.46}^{+0.47}$ and $\epsilon_{\rm p}=85_{-10}^{+15}$ keV) to the April 28 event obtained by Konus-Wind \citep{konus_paper} is overplotted with the black dashed line. Spectra are normalized at $10$ keV in arbitrary units.
}
\label{fig:spectra}
\end{figure}

Despite the peculiar light curve and unusually hard spectra of the X-ray burst (\S \ref{s:FRB200428}), it is possible that there might be an underlying trapped FB, partially contributing to the thermal part of the entire X-ray burst spectrum.
Figure \ref{fig:spectra} compares the best-fit model of observed hard X-ray spectrum \citep{konus_paper} with the predicted spectra from the trapped FB emission \citep{lyubarsky02,yamasaki20} with an effective temperature of $T_{\rm obs}\sim10$ keV, which is consistent with the black body (BB) plus power-law spectral fitting result with a temperature of $\sim11$ keV \citep{hxmt_paper}. One can clearly see that the thermal component of the observed spectra could be roughly described by these models (the excess in the non-thermal component will be discussed later in this section). Therefore, we assume that the thermal component of the X-ray burst spectrum might be interpreted as radiation from a trapped FB with a peak photon energy of $\epsilon_{\rm obs}\approx 3T_{\rm obs}\sim30$ keV, where a factor of $3$ reflects the Wien's displacement law. 

In the presence of a very strong magnetic field exceeding the critical quantum value $B_{\rm cr}\equiv m_e^2c^3/(e \hbar)\simeq4.4\times10^{13}$ G, the magnetic equilibrium pair number density is expressed as \citep{canuto77}
\beqn
\label{eq:n_eq_mag}
n_{\rm e, eq}^{\rm mag}(T)\approx\frac{1}{\sqrt{2\pi^3}}\,\lambda_{\rm C}^{-3}\,\frac{B}{B_{\rm cr}}\left(\frac{T}{m_ec^2}\right)^{1/2}\,e^{-m_ec^2/T},
\eeqn
where $\lambda_{\rm C}=\hbar/(m_ec)$ is the electron Compton wavelength and the numerical factor $(2\pi^3)^{-1/2}\lambda_C^{-3}\simeq 8.1\times10^{29}\ {\rm cm^{-3}}$. 
The energy transfer of the trapped FB under the strong magnetic field is governed by extraordinary-mode (X-mode) photons \citep{td95,lyubarsky02} with an effective Compton scattering cross section \citep{meszaros92}
\beqn
\label{eq:sigma_eff}
\sigma_{\rm eff}(\epsilon)=\sigma_{T}\left(\frac{\epsilon}{m_ec^2}\right)^2\left(\frac{B}{B_{\rm cr}}\right)^{-2},
\eeqn
where $\epsilon$ is the X-mode photon energy. The emergent spectrum is determined by the radiation spectrum at the depth corresponding to the mean free path of an X-mode photon
\beqn
l(T,\epsilon)\sim\frac{1}{n_{\rm e, eq}^{\rm mag}(T)\,\sigma_{\rm eff}(\epsilon)}.
\eeqn
By solving the energy transfer equations across the trapped FB, \citet{lyubarsky02} found that the emergent spectrum is well approximated by a modified BB with an effective temperature $T_{\rm obs}$. Because of the photon energy dependence of the X-mode cross-section (Eq. [\mbox{\ref{eq:sigma_eff}}]), only the Wein part of the BB spectrum with temperature $T_0$ is effectively observable as an outgoing radiation with temperature $T_{\rm obs}$, and therefore in general $T_0>T_{\rm obs}$. For simplicity, let us consider an effective X-mode photosphere with a radius $R_0$ and a uniform background FB temperature $T_0$, emitting a radiation with a photon energy of $\epsilon_{\rm obs} = 30$ keV. 
Assuming a uniform magnetic field of $B=B_{p}$, the mean free path for X-mode photons is $l(T_0,\epsilon_{\rm obs})\lesssim{\cal O}(1)$ cm for $T_0/m_ec^2\gtrsim0.1$, which is vanishingly small compared to the expected FB size $R_0$.

\begin{figure}
\centering
\includegraphics[width=.48\textwidth]{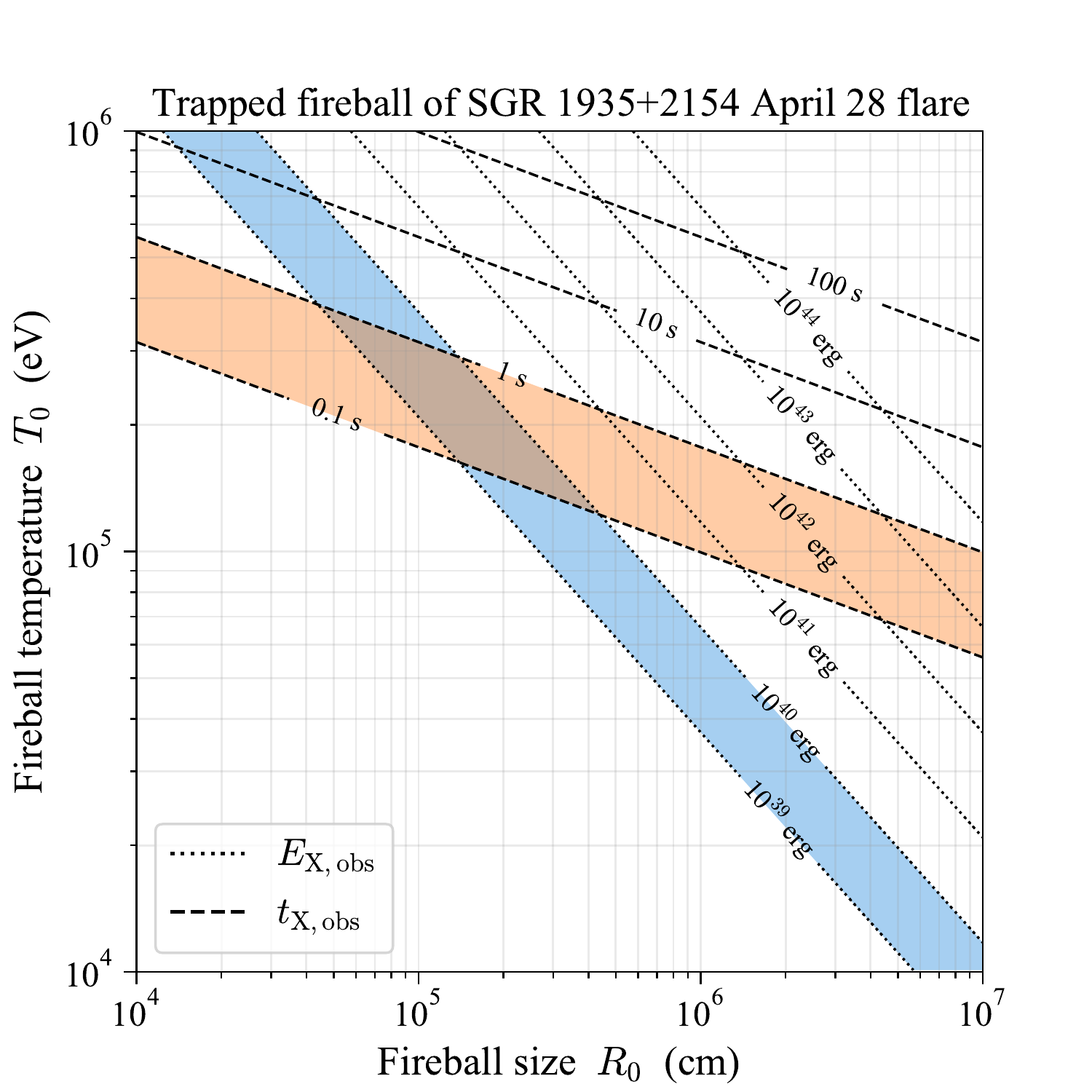}
\caption{The estimated radius versus temperature of trapped FB for SGR 1935$+$2154 (colored regions), assuming the observed photon energy of $\epsilon_{\rm obs}=30$ keV. The dotted and dashed lines represent the contours for the observed energy and duration, respectively.}
\label{fig:trappedFB}
\end{figure}

In the above simple picture \citep{td95,td96,lyubarsky02}, the total energy of the trapped FB is dominated by the hot plasma component with radius $R_0$ and temperature $T_0$ as
\beqn
\label{eq:E_trappedFB}
E_{\rm X,obs} =\frac{4}{3}\pi R_0^3\, a T_0^4,
\eeqn
where $a$ is the radiation constant. Meanwhile, the photon diffusion occurs only at the outermost surface with the observed luminosity
\beqn
\label{eq:L_fb}
L_{\rm X} =4\pi cR_0^2\,a T_{\rm obs}^4. 
\eeqn
Combining Eqs. \eqref{eq:E_trappedFB} and \eqref{eq:L_fb}, the observed duration of the X-ray emission from the trapped FB is estimated as
\beqn
t_{\rm X,obs}\sim\frac{E_{\rm X,obs}}{L_{\rm X}}.
\eeqn
Figure \ref{fig:trappedFB} shows the constrains on the trapped FB parameters. If we conservatively take $E_{\rm X,obs}=10^{39}$--$10^{40}$ erg and $t_{\rm X,obs}=0.1$--$1$ s, the allowed parameter space for the FB radius and temperature are $R_0\sim10^5$ cm and $T_0\sim200$--$300$ keV, respectively. With a BB temperature of $200$--$300$ keV, a good fraction ($70$\%--$87$\%) of the total energy is carried by photons with $\epsilon>m_ec^2$, and hence it is sufficient to keep the interior of trapped FB (except for the thin outer layer) optically-thick to pair production.
Given the short duration of the emission compared to the spin period  $t_{\rm X,obs}/P_{\rm spin}\lesssim 0.3$ as well as the relatively small FB size with respect to the NS, the FB may evaporate before being occulted due to the NS rotation.

As mentioned in \S \ref{s:FRB200428}, the spectrum of April 28 event is much
harder than that of typical bursts from SGR 1935+2154 with comparable duration and total energy. The resonant cyclotron scattering may be responsible for the spectral hardening inside flaring magnetosphere.
The magnetar magnetosphere is filled with $e^{\pm}$ plasma both during flares and in the persistent state \citep{tlk02,beloborodov07,beloborodov13}; one can easily see that the resonant cyclotron optical depth is unavoidably large. Therefore any outgoing radiation is reprocessed in the cyclotron resonance layer. 
In this case, a Doppler shift due to scattering on the bulk motions of the magnetospheric plasma could lead to formation of hard tails in thermal spectra. During the flare, a tremendous resonance radiation force keeps the plasma motion mildly relativistic \citep{yamasaki20}. As a result, under typical conditions for flaring magnetosphere, the degree of spectral hardening by a single scattering is at most twice in terms of observed photon energy and the single scattering model can successfully fit the observed intermediate flare (with  $L_{\rm X}\sim10^{40}$--$10^{41}\ {\rm erg\,s^{-1}}$) spectra from SGR 1900$+$14 (see, e.g., Figure 5 of \citealt{yamasaki20}).
Figure \ref{fig:spectra} clearly indicates that the predicted spectrum from the trapped FB emission reprocessed by a single resonant cyclotron scattering cannot explain the hard spectral index of April 28 event.
Hence, while most spectra of ordinary bursts from SGR 1935$+$2154 might be explained by this model, the formation of the extremely hard spectra of April 28 event by the same picture seems challenging unless one invokes an extremely dense magnetosphere that could lead to multiple resonant scatterings (see also \citealt{ioka20,yang21}). Since a further exploration of such a possibility is outside the scope of this work, we just note that magnetospheric reprocession of the trapped FB emission or some alternative mechanisms may give rise to the observed hard X-ray spikes in \S \ref{ss:constraints_trappedFB}.

\subsection{Relativistic Outflow}
\label{ss:outflow models}
Next we consider the relativistic outflows, which might be launched at the onset of the trapped FB formation and produce the radio burst and hard X-ray spikes.
The intrinsic energy budget for launching relativistic outflows should be limited by the isotropic equivalent energy emitted by the trapped FB ($E_{\rm X}^{\rm iso}=10^{40}$ erg).
Given the small variability timescale for radio and X-ray bursts ($\lesssim10$ ms), the maximum injected energy available for the outflow would be smaller than $E_{\rm X}^{\rm iso}$. Thus, we conservatively set the initial outflow energy to $E_{\rm flare}\sim10^{39}$ erg.
In addition to the energy and the initial size, the dynamical evolution of the outflow depends on both the composition of the FB and the energy source for acceleration, which are highly uncertain.
Thus, we consider a broad class of the outflow models in \S \ref{sss:Outflow Models}
and 
discuss its relevance to generation of coherent radio emission in \S \ref{sss:Emission Mechanisms}.

\subsubsection{Outflow Models}
\label{sss:Outflow Models}

We consider three outflow models: (i) leptonic outflow composed of $e^{\pm}$ pairs and photons, (ii) baryonic outflow composed of $e^{\pm}$ pairs, baryons and photons and (iii) magneto-leptonic (or simply, magnetic) outflow composed of cold $e^{\pm}$ pairs loaded with large Poynting flux.
We use the theory of an adiabatic FB \citep{paczynski86,goodman86} to track the dynamical evolution of these outflows (see Appendix \ref{s:appendix}). The evolution of leptonic outflow is uniquely determined for a given set of initial outflow parameters, such as size, temperature and bulk Lorentz factor (or energy), whereas the latter two outflows are characterized by additional model parameters.

The baryonic outflow is characterized by the baryon loading parameter $\eta$ defined as a ratio of radiation flux to matter energy flux. The magnetic outflow is described by means of its initial magnetization parameter $\sigma_0$ defined as a ratio of Poynting flux to matter energy flux at the magnetosonic point where the outflow attains a velocity of $\Gamma_0=\sigma_0^{1/2}$ and starts to evolve further\footnote{Note that the definition of $\sigma_0$ here is different from the conventional one that defines it when the flow is static ($\Gamma_0=1$).} (see Appendix for further details). In order to accelerate the magnetic outflow efficiently, a strong dissipation may be important\footnote{One caveat of models with strong dissipation, however, may be that it is not clear whether a highly ordered magnetic field can be maintained at the FRB generation site in order for the synchrotron maser mechanism to operate.}.
We adopt a classic model proposed by \citealt{drenkhahn02} in the context of gamma-ray bursts (GRBs), in which the toroidal magnetic field with alternating polarity (so-called striped wind model; \citealt{kc84,lyubarsky01}) decays into kinetic energy above the light cylinder [$r_{\rm lc}=cP_{\rm spin}/(2\pi)\sim10^{10}$ cm for SGR 1935$+$2154]. With the assumption that the outflow is highly dominated by magnetic energy and that the thermal energy is negligible we derive the dynamical evolution at $r>r_{\rm lc}$. 
We adopt a classic model proposed by \citealt{drenkhahn02} in the context of gamma-ray bursts (GRBs), in which the toroidal magnetic field with alternating polarity (so-called striped wind model; \citealt{kc84,lyubarsky01}) decays into kinetic energy above the light cylinder [$r_{\rm lc}=cP_{\rm spin}/(2\pi)\sim10^{10}$ cm for SGR 1935$+$2154]. With the assumption that the outflow is highly dominated by magnetic energy and that the thermal energy is negligible we derive the dynamical evolution at $r>r_{\rm lc}$. 

Based on the trapped FB properties estimated in \S \ref{ss:trappedFB}, we fix the initial non-magnetic outflow radius and temperature to be $r_0=R_0\sim10^5$ cm and $T_0\sim 200$ keV respectively, so that $E_{\rm flare}=4/3\pi r_0^3 aT_0^4$. The initial density of non-magnetic outflow is set to the thermal equilibrium value, which only depends on $T_{0}$. Meanwhile, the evolution of the magnetic outflow is calculated from $r=r_{\rm lc}$. Since we use the cold approximation, the initial density is determined by equating the initial kinetic energy to $E_{\rm flare}$. The black curves in Figures \ref{fig:leptonicFB}--\ref{fig:magneticFB} show the dynamical evolution of each outflow. The evolution of leptonic outflow is uniquely determined, whereas for baryonic and magnetic outflows we show the evolution with  characteristic values of $\eta$ and $\sigma_0$.To summarize, the terminal bulk Lorentz factor that each outflow attains is
\beqn
\label{eq:Gamma_infty}
    \Gamma_{\infty}\sim \begin{cases}
    4.4\times10^2\ r_{0,5}^{1/4} \,\hat{\Theta}_0& ({\rm Leptonic}) \\
    {\rm min}[\eta,\eta_{\rm heavy}] & ({\rm Baryonic})\\
    \sigma_0^{3/2}+\sigma_0^{1/2} & ({\rm Magnetic}),
  \end{cases}
\eeqn
where $\Theta_0=T_0/m_ec^2$ is the dimensionless initial FB (outflow) temperature and hereafter we use a notation $\hat{\Theta}_0\equiv\Theta_0/0.4$, corresponding to $T_0=200$ keV. We cover the baryonic outflows in heavy ($\eta<\eta_{\rm heavy}$) and mild ($\eta_{\rm heavy}<\eta<\eta_{\rm mild}$) load regimes, where $\eta_{\rm heavy}\sim60\,r_{0,5}^{1/4}\,\hat{\Theta}_0$ and $\eta_{\rm mild}\sim1.5\times10^4\,\hat{\Theta}_0\, r_{0,5}$ are the critical values (see Appendix \ref{s:appendix}).

One concern regarding the early evolution of the outflow is the possible disturbance by the large-scale magnetic field of the magnetar. Given a dipole magnetic field $B\propto r^{-3},$
the background magnetic pressure at an altitude $h$ above the NS surface is
$P_{\rm B}= B^2/(8\pi)\sim4\times10^{26} \ B_{\rm p,14}^2\,h_{6}^{-6}\ {\rm erg \ cm^{-3}}$, whereas the total pressure of the non-magnetic outflow with initial temperature $T_0$ is
$P_{\rm fb}=aT_0^4\sim3\times10^{23}\,\hat{\Theta}_0^4\ {\rm erg \ cm^{-3}}$. Namely, $P_{\rm B}\gtrsim P_{\rm fb}$ at an altitude $h\lesssim h_{\rm c}\sim3\times10^{6}\ B_{\rm p,14}^{1/3}\,\hat{\Theta}_0^{-2/3}$ cm. While a leptonic outflow is barely affected by the background magnetic field because it continues to accelerate up to much larger distance $r_{\infty}=\Gamma_{\infty}r_0\sim4.4\times10^7\ r_{0,5}^{5/4} \,\hat{\Theta}_0$ cm compared to $h_{\rm c}$,
it may significantly modify the early evolution of baryonic outflows with low acceleration efficiency  $r_{\infty}\lesssim \eta_{\rm heavy}r_0\sim6.0\times10^6\ r_{0,5}^{5/4} \,\hat{\Theta}_0$ cm, which is almost comparable to $h_{\rm c}$. In this respect, our estimate on $\Gamma_{\infty}$ could be slightly overestimated. 
The situation might be more complicated for cold magneto-leptonic outflows due to the absence of the radiation pressure. Nevertheless, such uncertainties must be sub-dominant relative to the assumption that the flow starts to evolve at $r=r_{\rm lc}$ with significant acceleration $\Gamma_0=\sigma_0^{1/2}$. Therefore, we neglect the potential modification of inner outflow evolution by the background magnetic field hereafter.

\subsubsection{Plasma cutoff frequency}
\label{sss:Emission Mechanisms}

It is often assumed that the GHz coherent emission is generated by coherent charge bunches through, e.g., curvature or synchrotron maser
processes. 
In the case of curvature radiation, the emission is often thought to be triggered by magnetic reconnection in the vicinity of NS \citep[e.g.,][]{katz16,kumar17,ghisellini18,lu18,yang18,katz20,lu20}.
In the case of the synchrotron maser emission, the emission occurs at relativistic shocks propagating in the pre-existing media, such as  nebula \citep{lyubarsky14,murase16,waxman2017}, steady magnetar wind \citep{beloborodov17}, or past flare-driven ejecta \citep{metzger19,beloborodov20,margalit20,yuan20,yu20}.

In either case, one of the important constraints for localizing the radio emission region comes from the plasma cutoff effect.
The waves have cutoff frequencies $\omega_{\rm cutoff}$ (measured in the plasma frame) below which they become evanescent.
In general, the cut-off frequency is conveniently expressed in terms of the plasma frequency $\omega_{\rm p}$ defined in the plasma rest frame as
\beqn
\label{eq:nu_p_general}
\omega_{\rm p}\equiv\zeta\sqrt{\frac{4\pi n_e^{\prime} e^2}{m_e}},
\eeqn
where $n_e^{\prime}$ is the comoving number density of electrons in region which is responsible for the generation of waves. We include all the uncertainties associated radiation mechanisms and plasma conditions in the fudge factor $\zeta$, representing both the possible relativistic effects and the specific treatment of the shock.  
Throughout this work, for simplicity, we set $\zeta=1$ and leave the parameter dependence to keep generality.

In the case of maser emission at far zone, electromagnetic (EM) waves follow the well-known dispersion relation in the non-magnetized plasma with a cutoff at $\omega_{\rm cutoff}=\omega_{\rm p}$\footnote{In the case of curvature process near the NS, the cyclotron frequency of electrons or positrons $\omega_{\rm B}=eB/(2\pi m_{\rm e}c)$, where $B$ is the local magnetic field strength, is typically much greater than the wave frequency and/or local plasma frequency. In this case (see also \S \ref{ss:trappedFB}), there are two polarization states of EM waves (O-mode and X-mode).  
While the O-mode wave has the same dispersion relation as in the non-magnetized plasma with a cutoff at $\omega_{\rm cutoff}=\omega_{\rm p}$, the X-mode wave has a complicated dispersion relation with two cutoffs. The lower cutoff lies at $\omega_{\rm cutoff}=(\omega_{\rm p}^{2}+\omega_{\rm B}^2/4)^{1/2}-\omega_{\rm B}/2\sim\omega_{\rm p}^{2}/\omega_{\rm B}$ when $\omega_{\rm p}\ll\omega_{\rm B}$ (e.g., \citealt{chen84,arons86}), indicating $\omega_{\rm cutoff}\ll\omega_{\rm p}$. Depending on how much fraction of the radiation is in X-mode, the condition for the wave propagation may be much more relaxed compared to the non-magnetized plasma case \citep{kumar17}. By incorporating this effect, one may estimate the apparent plasma frequency in the observer frame for the curvature-type scenario as
$\omega_{\rm p,obs}=\Gamma\omega_{\rm p}\,{\rm min}[(\omega_{\rm obs}^{\prime}/ \omega_{\rm B})^{1/2},1]$, where $\omega_{\rm p}$ is estimated by Eq. \eqref{eq:nu_p_general}.}, but the treatment of the shocked region becomes important for an appropriate estimate of plasma frequency.
As seen in \S \ref{sss:Outflow Models}, we calculate the dynamical evolution of a single outflow ($\Gamma$ and $n_e^{\prime}$) without deceleration, which may differ from the exact quantitative dynamics of decelerating outflow shells that produce internal shocks.
Nevertheless, we can reasonably assume that the most efficient internal shock with a large contrast between shell Lorentz factors and comparable densities is generated at each radius $r$. 
We assume that the upstream (downstream) of the shock is cold (hot), and the maser emission is produced by the cold upstream plasma at the shock front. In this case, the apparent plasma frequency in the observer frame for maser-type scenarios is evaluated by (\citealt{plotnikov19}, see also \citealt{iwamoto17,iwamoto19})
\beqn
\label{eq:nu_p_maser}
\omega_{\rm p,obs}\approx \Gamma\omega_{\rm p}\,\rm max[1,\sigma^{1/2}],
\eeqn
where the coefficient of $3$ appearing in the original formula is neglected for simplicity. Here, again, there is an uncertainty in the treatment of bulk Lorentz factor depending on the shock models. But, as this is small $\sim1$, it can be absorbed by the fudge factor $\zeta$ in Eq. \eqref{eq:nu_p_general}.

\section{Constraints on Relativistic Outflow and Emission Region}
\label{s:outflow constraints}

Based on the outflow models outlined in \S \ref{ss:outflow models}, we aim to obtain general constraints on the properties of the outflow that is responsible for the generation of radio and hard X-ray bursts from SGR 1935$+$2154. 

\subsection{Coherent Radio Burst}
\label{ss:constraints_trappedFB}

Radio emission suffers from various constraints when escaping from the system without significant attenuation, and there is a radio compactness problem when the radio emission originates from relativistic outflows. For example, \cite{murase17} investigated whether radio emission can coincide in region with X-ray and gamma-ray emission in light of FRB 131104 \citep{2016ApJ...832L...1D}. Radio waves can propagate only when their frequencies are higher than the plasma cutoff frequency and they also suffer from the induced Compton scattering within the outflows and ambient environments \citep[e.g.,][]{murase16}. Here we focus on the plasma cutoff condition for the radio wave propagation:
\beqn
\label{eq:plasma_cutoff}
\omega_{\rm p,obs}(r_{\rm radio})\lesssim\omega_{\rm obs},
\eeqn
where $\omega_{\rm p,obs}=\omega_{\rm p}$ is the apparent plasma frequency in the observer frame. We set the observed radio frequency to $\nu_{\rm obs}=\omega_{\rm obs}/(2\pi)=1$ GHz in mind of CHIME and STARE2. 
Depending on the radial evolution of the observed plasma frequency, the above condition sets a limit on the radio-emitting radius $r_{\rm radio}$. 

Another constraint comes from the intrinsic timing of radio and X-ray bursts. When there is a bulk motion with a Lorentz factor of $\Gamma$, the comoving size of the region responsible for the generation of emission can be larger by a factor of $\Gamma^2$. Given the intrinsic time delay $\Delta t_{\rm X, radio}\lesssim10$ ms (see \S \ref{s:FRB200428}), the radio (or X-ray) photons should be emitted at 
\beqn
\label{eq:time_delay}
r_{\rm radio \,(X)}\lesssim\Gamma^2 c\Delta t_{\rm X, radio}
\eeqn
which gives an upper limit on the radio (or X-ray) emitting radius. Since the time delay between X-ray and radio emissions generally depends on the emission mechanisms and initial FB size, it could be much shorter. Also, when there is little or no time delay between X-ray and radio emission as predicted by some models (e.g., \citealt{metzger19,margalit20,yuan20}), the time delay argument (Eq. [\ref{eq:time_delay}]) could be less constraining.
In this sense, the above limit is most conservative.

Given relativistic outflow models (\S \ref{sss:Outflow Models}) and maser-type emission (\S \ref{sss:Emission Mechanisms}), the plasma frequency argument (Eq. [\ref{eq:plasma_cutoff}]) sets the lower limit on the radio-emitting radius
\beqn
\label{eq:plasma_cutoff_}
&&r_{\rm radio}\gtrsim r_{\rm cutoff}\\
&&\sim\begin{cases}\nonumber
    3.7\times10^{13}\ r_{0,5}^{5/8}\,\hat{\Theta}_0^2\,\zeta\, \nu_{\rm obs,9}^{-1} \ \ {\rm cm},& ({\rm L}) \\
   1.1\times10^{13}\ r_{0,5}\,\hat{\Theta}_0^2\,\zeta\, \nu_{\rm obs,9}^{-1}\,{\rm min}[\left(\eta/\eta_{\rm heavy}\right)^{-1/2},1]\ \,{\rm cm} & ({\rm B})\\
    2.9\times10^{14}\ r_{0,5}^{-1/2}\,E_{\rm flare,39}^{1/2}\,\zeta\,\nu_{\rm obs,9}^{-1}\ \,{\rm cm} & ({\rm M}),
\end{cases}
\eeqn
where $r_{\rm cutoff}$ is the plasma cutoff radius defined by $\nu_{\rm p,obs}(r_{\rm cutoff})=\nu_{\rm obs}$. 
Next, the time delay argument (Eq. [\ref{eq:time_delay}]) suggests that the radio emission be emitted at
\beqn
\label{eq:time_delay_}
&&r_{\rm radio}\lesssim\Gamma_{\infty}^2c\Delta t_{\rm X, radio}\\
&&\sim\left(\frac{\Delta t_{\rm X, radio}}{10 \ \rm ms}\right)\times\begin{cases}\nonumber
    5.8\times10^{13}\ r_{0,5}^{1/2}\,\hat{\Theta}_0^2\, \ {\rm cm},& ({\rm L}) \\
   3.0\times10^{14}\ {\rm min}[\eta_{3}^2,\eta_{\rm heavy,3}^2] \, \ {\rm cm}, & ({\rm B})\\
   3.0\times10^{14}\ \sigma_{0,2}^{3}\,\ {\rm cm}, & ({\rm M}),
\end{cases}
\eeqn
which gives an upper limit on the radio-emitting radius. Here we set $\Gamma=\Gamma_{\infty}$ to make the radial constraints most conservative. 

\begin{figure}
\centering
\includegraphics[width=.5\textwidth]{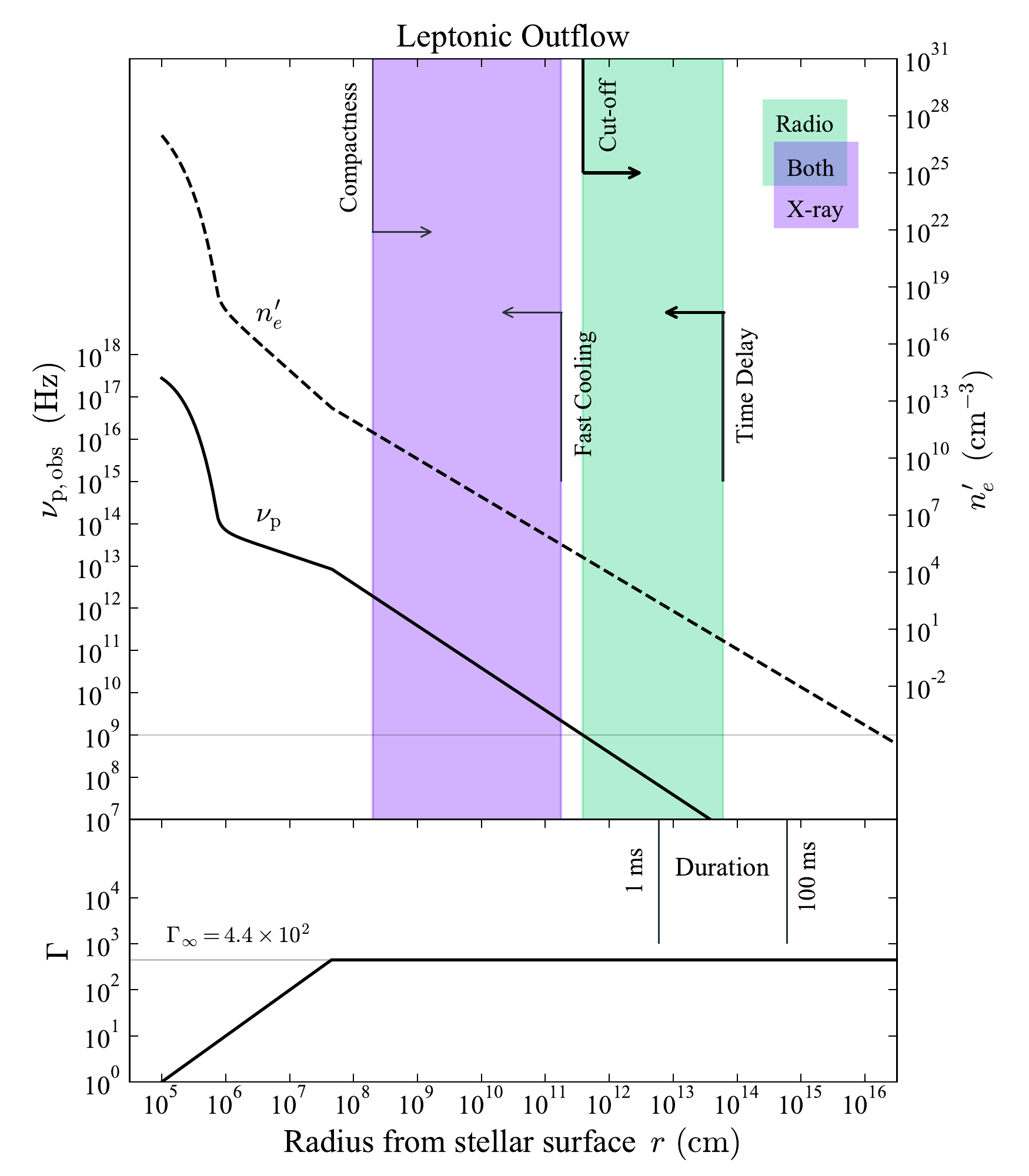}
\caption{Dynamical evolution of electron number density in the plasma rest frame (upper panel, right-hand-side axis), plasma frequency in the observer frame (upper panel, left-hand-side axis) and bulk Lorentz factor (lower panel) of the leptonic outflow with $E_{\rm flare}=10^{39}$ erg ($r_0=10^5$ cm and $T_0=200$ keV). The allowed radii for X-ray and radio emission are indicated by shaded regions in the upper panel. The region corresponding to the observed burst duration of $1$--$100$ ms is indicated by vertical lines in the lower panel. For radio emission Eq. \eqref{eq:plasma_cutoff_} and Eq. \eqref{eq:time_delay_} are used while for X-ray emission Eq. \eqref{eq:compactness} and Eq. \eqref{eq:cooling} are used. We assume $\Delta t_{\rm X,radio}=10$ ms when deriving the time delay constraints and the duration is evaluated by means of Eq. \eqref{eq:duration}.}
\label{fig:leptonicFB}
\end{figure}

\begin{figure*}
\begin{minipage}{0.49\hsize}
  \begin{center}
   \includegraphics[width=1.02\textwidth]{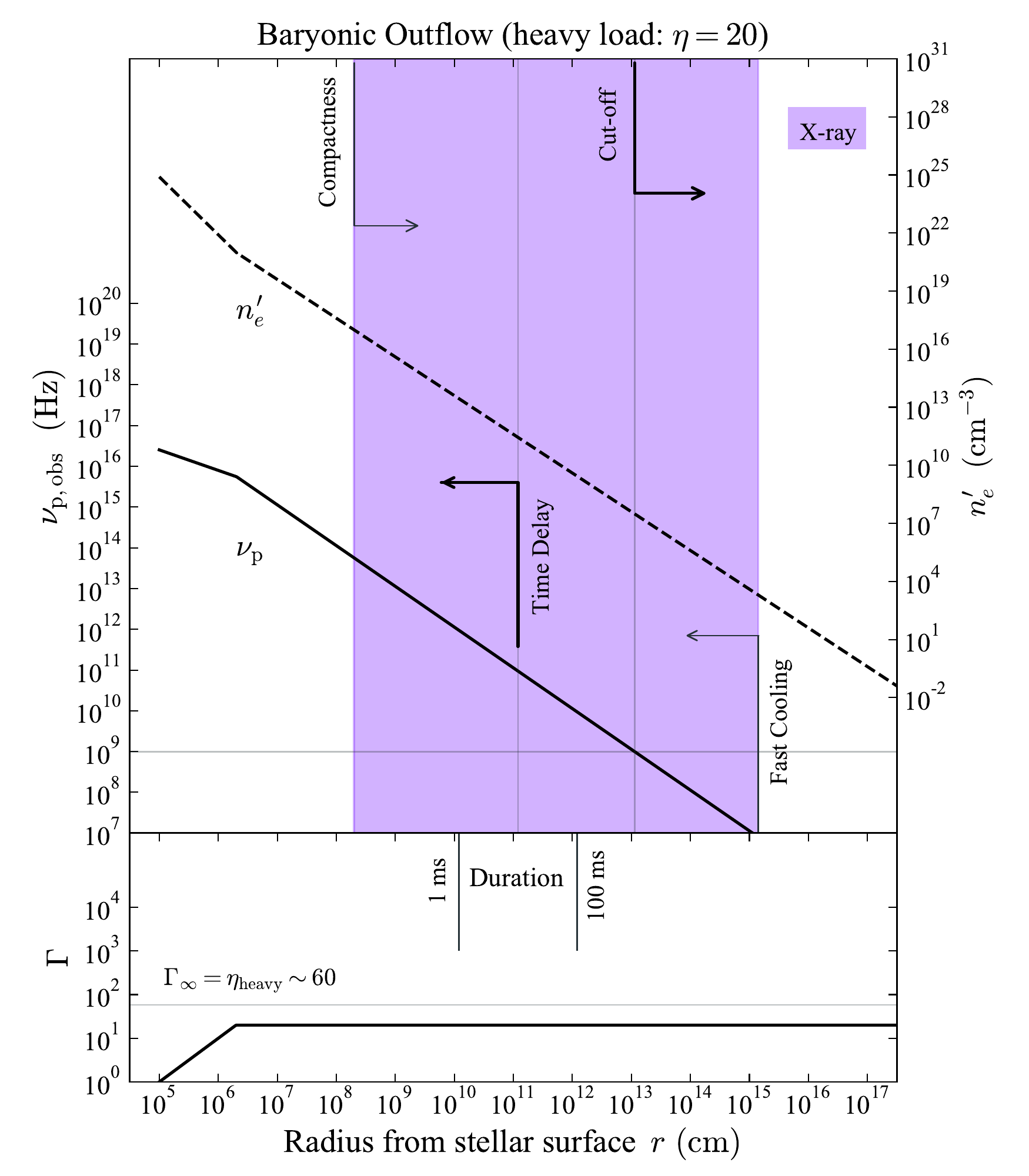}
  \end{center}
 \end{minipage}
 \begin{minipage}{0.49\hsize}
  \begin{center}
   \includegraphics[width=1.02\textwidth]{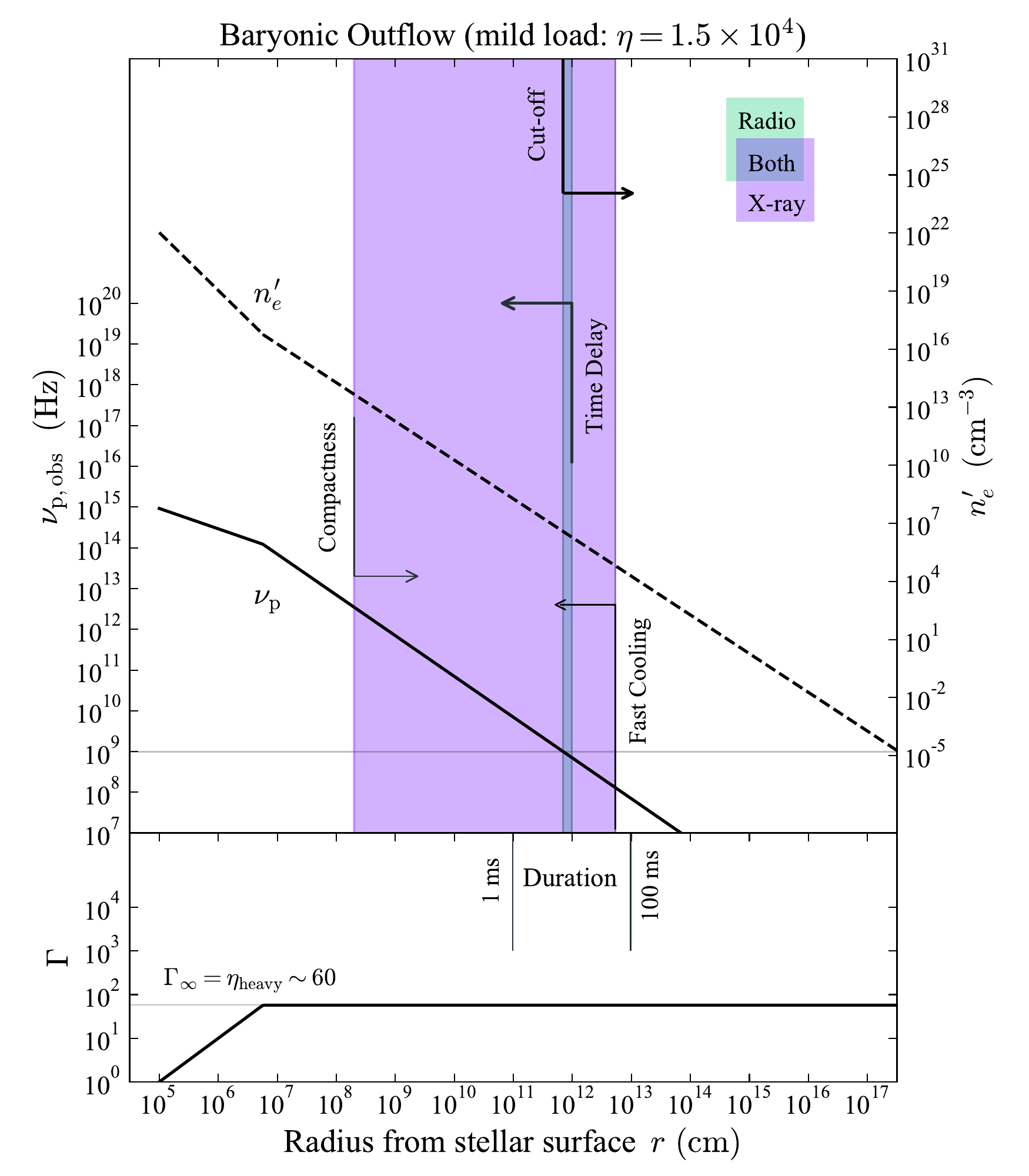}
  \end{center}
 \end{minipage}
 \caption{Same as Figure \ref{fig:leptonicFB} but for baryonic outflows in the heavy-load $\eta<\eta_{\rm heavy}$ (left) and mild-load $\eta_{\rm heavy}<\eta<\eta_{\rm mild}$ (right) regimes. In the limit of extremely weak baryon load $\eta\to\infty$, the dynamical evolution of the outflow asymptotically approaches to that of a pure leptonic one shown in Figure \ref{fig:leptonicFB}. Note that the upper limit on the radii due to the fast cooling scales as $\propto\epsilon_B$ (Eq. [\ref{eq:cooling}]) and could be much smaller than shown here (we take an extreme limit $\epsilon_B=1$), in which case the radio-emitting region may not overlap with the X-ray-emitting region.}
\label{fig:baryonicFB}
\end{figure*}

\begin{figure*}
\begin{minipage}{0.49\hsize}
  \begin{center}
   \includegraphics[width=1.02\textwidth]{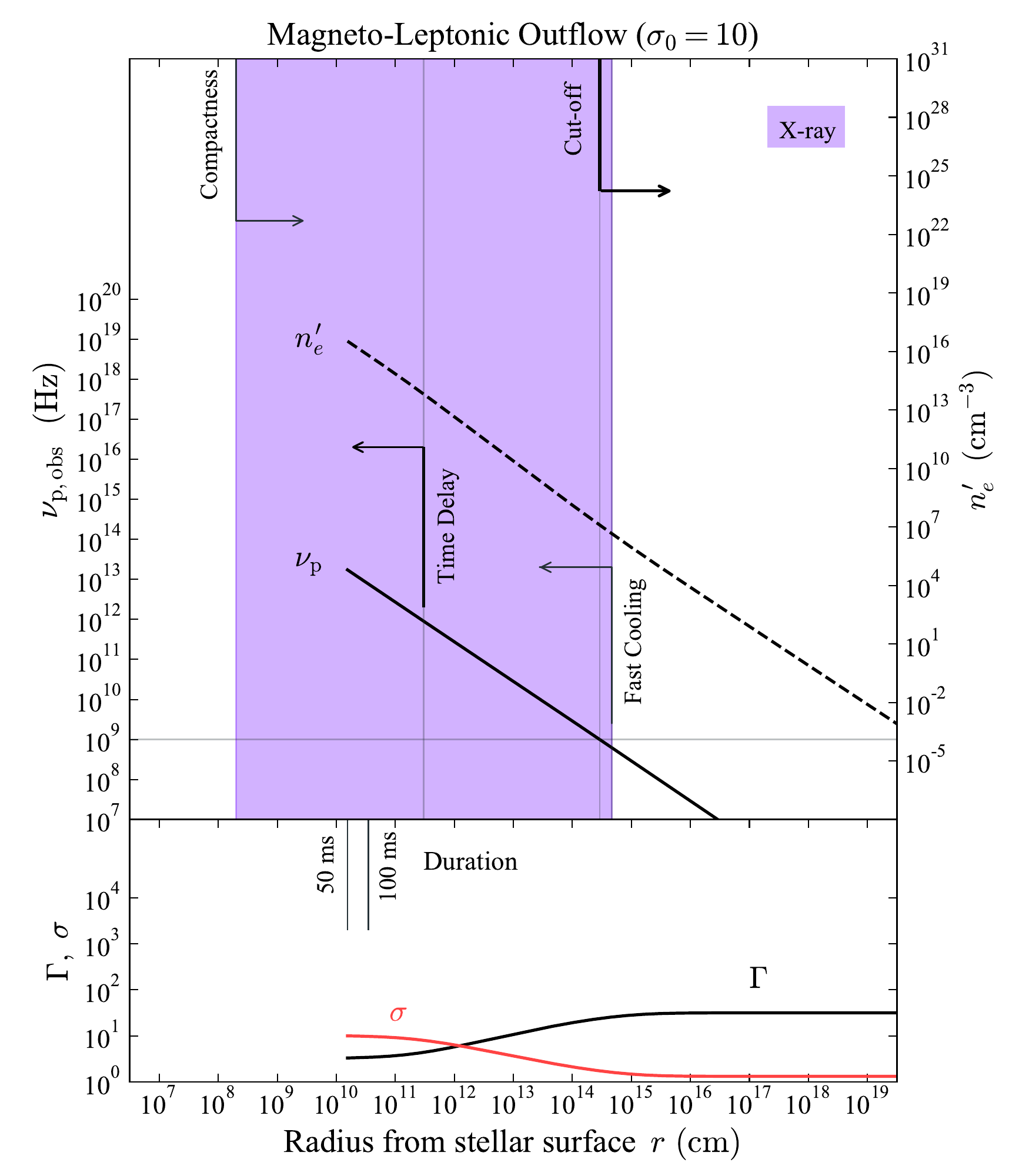}
  \end{center}
 \end{minipage}
 \begin{minipage}{0.49\hsize}
  \begin{center}
   \includegraphics[width=1.02\textwidth]{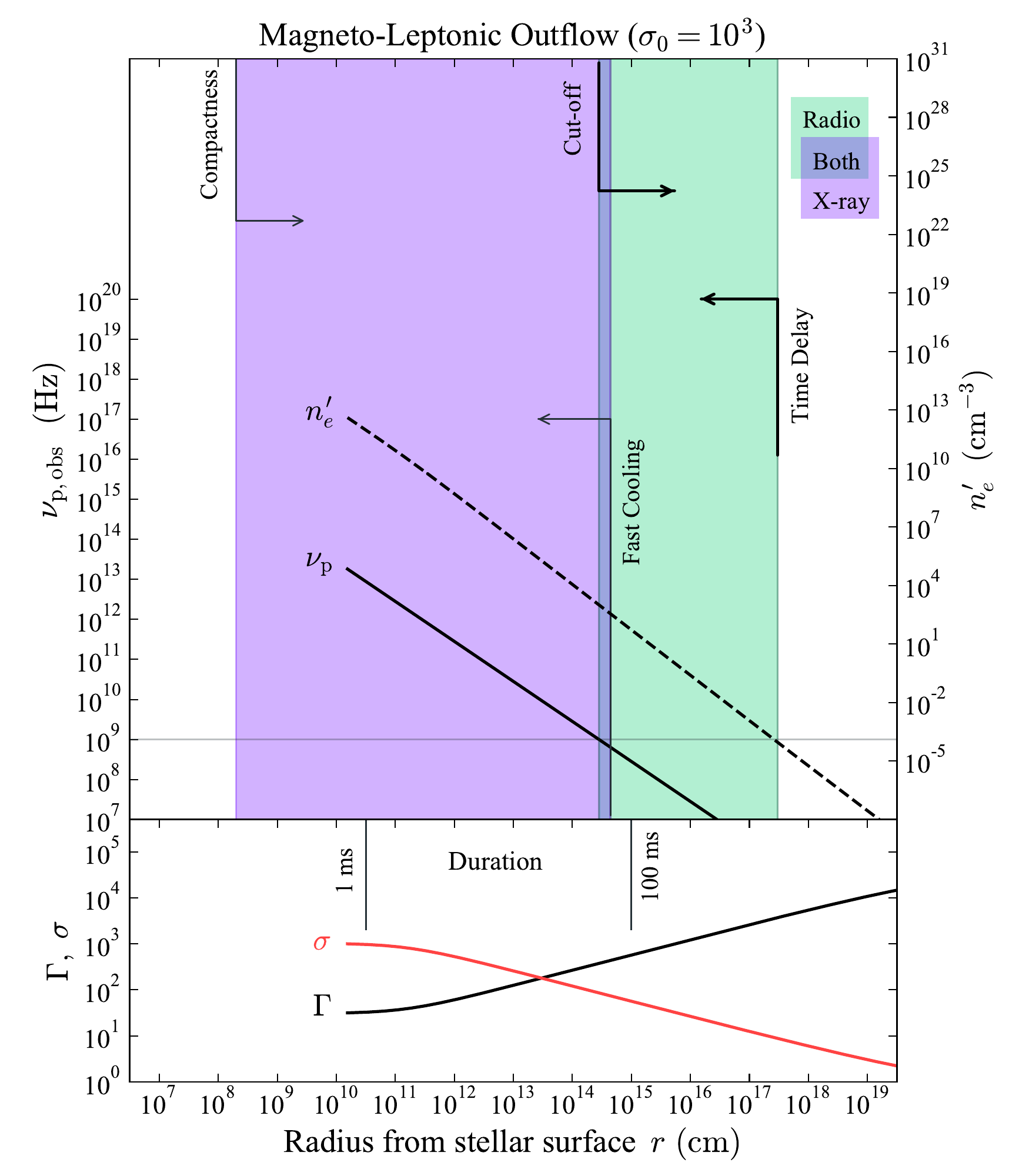}
  \end{center}
 \end{minipage}
 \caption{Same as Figure \ref{fig:leptonicFB} and \ref{fig:baryonicFB}  but for magneto-leptonic outflows with initial degree of magnetizations $\sigma_0=10$ (left) and $\sigma_0=10^3$ (right). The outflow energy is set to $E_{\rm flare}=10^{39}$ erg at $r=r_{\rm lc}$. The radial evolution of magnetization parameter $\sigma$ is also shown in the lower panels. }
\label{fig:magneticFB}
\end{figure*}

The allowed region for the radio emission, as well as dynamical evolution, are indicated by the vertical green shaded regions in Figures \ref{fig:leptonicFB}--\ref{fig:magneticFB}. One can see from Figure \ref{fig:leptonicFB} that the allowed locale of radio emission from a leptonic outflow is constrained to within somewhat narrow regions at $r_{\rm radio}\sim10^{12}$--$10^{14}\ {\rm cm}$.
On the other hand, the evolution of bulk Lorentz factor of baryonic and magnetic outflows strongly depends on the initial degree of baryon laod ($\eta$) and magnetization ($\sigma_0$). Figures \ref{fig:baryonicFB} and \ref{fig:magneticFB} demonstrate how these parameters affect the conditions of Eq. \eqref{eq:plasma_cutoff_} and Eq. \eqref{eq:time_delay_}. From the left panels of Figures \ref{fig:baryonicFB} and \ref{fig:magneticFB}, it is apparent that
heavily baryon-loaded and weakly magnetised outflows 
are not compatible with observed time delay due to the modest 
acceleration. Meanwhile, although the maximum acceleration is also limited in the the mild-load regime ($\eta_{\rm heavy}<\eta<\eta_{\rm mild}$; the left panel of Figure \ref{fig:baryonicFB}), there is an allowed range for radio emission site because of smaller plasma frequency. Similarly, the higher initial magnetization $\sigma_0$ results in the faster acceleration, which broadens the allowed range of emission region. 
As a consequence, the initial properties
\beqn
\label{eq:eta_limit}
&\eta& \gtrsim 6.2\times10^{3} \ r_{0,5}^{5/4}\,\hat{\Theta}_0\,\zeta^2\,\nu_{\rm obs,9}^{-2}\left(\frac{\Delta t_{\rm X,radio}}{\rm 10\ {\rm ms}}\right)^{-2}\\
\label{eq:sigma_limit}
&\sigma_0& \gtrsim 99\ r_{0,5}^{-1/6}\,E_{\rm flare,39}^{1/6}\,\zeta^{1/3}\,\nu_{\rm obs,9}^{-1/3} \left(\frac{\Delta t_{\rm X,radio}}{\rm 10\ {\rm ms}}\right)^{-1/3}
\eeqn
are required for each outflow to keep the consistency with arguments on the plasma cutoff frequency and the observed time delays between X-ray and radio emission, i.e., $r_{\rm cutoff}\lesssim\Gamma_{\infty}^2c\Delta t_{\rm X, radio}$. In Eq. \eqref{eq:sigma_limit},  we use an approximation $\Gamma_{\infty}=\sigma_0^{3/2}+\sigma_0^{1/2}\sim\sigma_0^{3/2}$ for simplicity. 

The observed duration of the burst emission that each outflow predicts can be estimated by
\beqn
\label{eq:duration}
\delta t\sim \frac{r}{c\Gamma^2}.
\eeqn
Considering the observed duration of interest $\delta t =1$--$100$ ms, we show the allowed emission region in the lower panels of Figures \ref{fig:leptonicFB}--\ref{fig:magneticFB}. By examining whether it overlaps with the radio-emitting region, one finds that the leptonic outflow, mildly-loaded baryonic outflow are in principle compatible with radio observations. Meanwhile, heavily baryon-loaded  and magnetized outflows cannot reproduce sufficiently short duration of radio bursts due to their weak or delayed acceleration. 

\subsection{Hard X-ray Bursts from Relativistic Outflow}
\label{ss:hard X-ray from outflows}

The existence of non-thermal component in the observed X-ray spectra implies that the source is optically thin to Thomson scattering on $e^{\pm}$ pairs\footnote{Since the maximum observed photon energy $\sim250$ keV \citep{konus_paper} of the hard X-ray counterpart to the radio burst on April 28 is well below $m_ec^2$, the opacity to $\gamma\gamma$ pair production provides less stringent constraints on the beaming of the outflow.}, which is often the case with the prompt emission of GRBs.
An inevitable source for such pairs is the annihilation of photons with rest-frame energy above $m_ec^2$.
The scattering optical depth for created pairs is expressed as \citep{nakar07,matsumoto19}
\beqn
\label{eq:tau_T}
\tau_{\rm T}\approx\frac{\sigma_{\rm T}f_{\rm th}N^{\prime}}{\pi\theta_{X}^2 r_{\rm X}^2},
\eeqn
where $N^{\prime}$ is the total number of emitted photons in the rest frame of the outflow and $f_{\rm th}$ the fraction of photons that create pairs. We approximate the energy-averaged cross section as $\sigma_{\rm T}$ for simplicity. 
The observer-frame quantities, $\theta_{X}$ and $r_{\rm X}$, are the geometric opening angle (relative to the outflow direction of motion) within which most of the photons propagate and radial distance of X-ray emission region, respectively.
The total number of photons is related to observed quantities by $N^{\prime}\approx (L_{\rm X}^{\rm iso}\delta t_{\rm X}/\epsilon_{\rm p})/\delta_{\rm D}^2(\theta,\Gamma)$,
where $L_{\rm X}^{\rm iso}$, $\delta t_{\rm X}$, $\epsilon_{\rm p}$ are the isotropic equivalent X-ray luminosity, the variability timescale (corresponding to the observed peak width of X-ray burst spikes), and the peak energy of photons in observed $\nu F_{\nu}$ spectra, respectively.
Here $\delta_{\rm D}\equiv1/[\Gamma(1-\beta\cos\theta)]$ denotes the Doppler factor corresponding to a Lorentz factor $\Gamma$ (and velocity $\beta$) and observer viewing angle $\theta$, which is measured from the center of the X-ray beam. The angular variation of the Doppler factor depends on the
product $\Gamma\theta$ and the size of X-ray emission region satisfies:
$\theta_{\rm X}\sim {\rm max}\left(1/\Gamma,\theta\right)$. Generally, the radial distance of X-ray emission region $r_{\rm X}$ is limited by the variability timescale $\delta t_{\rm X}$. Here we conservatively assume 
\beqn
\label{eq:r_X}
r_{\rm X}\sim \Gamma\beta\delta_{\rm D}(\theta,\Gamma)c\delta t_{\rm X},
\eeqn
which is true at least inside the
beam with angle $1/\Gamma$ regardless of specific dissipation mechanisms \citep{piran99} and indeed gives the loosest limit on the pair creation optical depth even outside the beam \citep{matsumoto19}. 
The relativistic beaming effect can also significantly change the pair-creation criteria and we define the energy threshold
of photons which can self-annihilate as $\epsilon_{\rm th}=\delta_{\rm D}(\theta,\Gamma)m_e c^2$ \citep{lithwick01}.
Then, the number fraction of annihilating photons in Eq. \eqref{eq:tau_T} is estimated by
\beqn
f_{\rm th}=\int_{\epsilon_{\rm th}}^{\infty}\frac{dN}{d\epsilon}d\epsilon,
\eeqn
where $dN/d\epsilon$ is the observed photon flux normalized to unity. The hard X-ray spectrum of FRB 200428 extends up to $250$ keV and is fitted by an exponentially-cutoff power law function $dN/d\epsilon\propto\epsilon^{\alpha}\exp{\left[-(\alpha+2)(\epsilon/\epsilon_{\rm p})\right]}$ with $\alpha=-0.72_{-0.46}^{+0.47}$ and $\epsilon_{\rm p}=85_{-10}^{+15}$ keV \citep{konus_paper}. 
Additionally we take $L_{\rm X}^{\rm iso}\sim10^{41}\ {\rm erg\ s^{-1}}$ and $\delta t_{\rm X}\sim10$ ms for the hard X-ray burst, so that isotropic energy is consistent with the total outflow energy $E_{\rm flare}\sim10^{39}$ erg.

Then, the requirement that $\tau_{\rm T}<1$ leads to the limit on observer viewing angle $\theta$, Lorentz factor $\Gamma$, and the radial distance $r_{\rm X}$ at which the X-ray emission escapes from the relativistic outflow. 
We find that the resulting constraints on the Lorentz factor and beaming are rather weak: $\theta\lesssim0.8$ and $\Gamma\gtrsim1$, which is largely due to the much lower peak energy and luminosity with respect to those of GRBs. Nevertheless, one can set a generic limit on the radius above which non-thermal emission can be produced as 
\beqn
\label{eq:compactness}
r_{\rm X}\gtrsim2\times10^8\ L_{\rm X,41}^{\rm iso 1/2} \delta t_{\rm X,-2}^{1/2}\ {\rm cm},
\eeqn
which is independent of outflow models presented in \S \ref{ss:outflow models}.

Provided that the hard X-ray burst is synchroton emission, the large flux of X-rays may ensure that X-ray emitting electrons would be in fast cooling regime regardless of its origins. For non-magnetic outflows, we assume that a fraction $\epsilon_B$ of the total internal energy density of the outflow is converted into magnetic energy in the frame of shocked fluid as $B^{\prime 2}\approx8\pi\epsilon_B \Gamma^2U^{\prime}$, where $U^{\prime}= m_e n_{e}^{\prime} c^2$ is the internal energy density of upstream material. For magnetic outflow, we can directly determine the magnetic field  behind the shock as $B^{\prime 2}\approx4\pi\sigma\Gamma^2U^{\prime}$. The synchrotron cooling Lorentz factor of outflow material is given by \citep{sari98}
\beqn
\label{eq:gamma_c}
\gamma_c = \frac{6\pi m_e c}{\sigma_T {B^{\prime}}^2 \Gamma \,t}, 
\eeqn
where $t\sim r/(\Gamma^2c)$ is the dynamical timescale of the flow in the frame of the observer. The typical Lorentz factor of electrons at the internal shock may be estimated by assuming that a fraction $\epsilon_e$ of the total internal energy goes into random motions of the electrons:
\beqn
\label{eq:gamma_m}
\gamma_{\rm m}\sim\epsilon_e \, \xi_e^{-1}\,\Gamma,
\eeqn
where $m_e/m_p\le\xi_e\le1$ is the fraction of electrons that undergo acceleration \citep{eichler05}. Here we take $\xi_e=1$, considering the maximum acceleration expected for an internal shock inside the (magneto-)leptonic outflow. Meanwhile, for baryonic outflow we choose $\xi_e=10^{-3}$, which may hold unless the flow is only weakly loaded with baryons ($\eta\gtrsim\eta_{\rm mild}\sim10^4$). Comparing the dynamical evolution of $\gamma_{\rm c}$ with $\gamma_{\rm m}$, one can show that the outflow is in fast-cooling regime ($\gamma_{\rm m}>\gamma_{\rm c}$) at
\beqn
\label{eq:cooling}
&&r_{\rm X}\lesssim\epsilon_e  \\
&&\times\begin{cases}\nonumber
    1.8\times10^{11} \ \,r_{0,5}^2 \,\hat{\Theta}_0^2\,\epsilon_{\rm B}\,\xi_e^{-1}\ {\rm cm}& ({\rm L}) \\
    1.4\times10^{15} \ \,r_{0,5}^2 \,\hat{\Theta}_0^4\,\epsilon_{\rm B}\,\xi_{e,-3}^{-1}\,{\rm min}[\left(\eta/\eta_{\rm heavy}\right)^{-1},1]\ {\rm cm} & ({\rm B})\\
    4.7\times10^{14} \ \,r_{0,5}^{-1}\,E_{\rm flare,39}\,\xi_e^{-1} \,\ {\rm cm} & ({\rm M}),
  \end{cases}
\eeqn
where we have used an analytic expression for the evolution of magnetic outflow (see Appendix \ref{s:appendix}). 
Hence, this could be considered as an upper limit on the X-ray emission radius. Clearly, the leptonic outflow cannot keep a high radiation efficiency far outside the magnetosphere. In Eq. \eqref{eq:cooling}, the possible uncertainty stemming from the treatment of bulk Lorentz factor used in Eqs. \eqref{eq:gamma_c} and \eqref{eq:gamma_m}, which depends on the detail of the shock model is neglected here.

By combining the available constraints on X-ray and radio emission with the duration constraint (Eq. [\ref{eq:duration}]), one finds that the leptonic outflow is excluded since it is unable to explain the X-ray burst duration. Due to the same reason, mildly-loaded baryonic outflows is also excluded. In contrast to the non-magnetic cases, high-$\sigma_0$ flows are marginally consistent with observations, albeit with somewhat long duration ($>10$ ms) for radio emission.

\section{Summary and Discussion}
\label{s:discussion}
In this work, we constrained the outflow properties associated with the unique April 28 event from SGR 1935$+$2154 consisting of radio and X-ray bursts. The event is likely to be triggered by sudden eruptions of magnetic energy of $\sim10^{39}$--$10^{40}$ erg into the magnetosphere, which would generate FB plasmas. 
As a consequence, a relativistic outflow might be launched at the onset of the trapped FB formation. 
In this case, the hard X-ray burst can be explained as a mixture of thermal and non-thermal emission. 
We showed that the thermal component of the X-ray burst spectrum is consistent with a trapped FB with temperature of a few hundred keV and size of $\sim10^5$ cm.

On the other hand, non-thermal radiation, including the non-thermal component of X-ray burst and the coherent radio burst, may arise from the relativistic outflow at large distances from NS ($r_{\rm X}\sim10^{8}$--$10^{10}$ cm and $r_{\rm radio}\gtrsim10^{11}$--$10^{12}$ cm) to avoid absorption/scattering by the outflow itself.
We calculated the dynamical evolution of the outflow so that its initial conditions are consistent with the inferred properties of the trapped FB. By assuming that these emissions are both produced by the energy dissipation at the internal shocks of the outflow, we show that any outflows should be accelerated up to bulk Lorentz factor of order $\sim10^2$ at the outer edge of magnetosphere. 

Furthermore, by examining the intrinsic timing offset between radio and X-ray burst spikes with $\lesssim 10$ ms, we constrain the initial degree of baryon load and magnetization, showing that $\eta\gtrsim6\times10^3$ and $\sigma_{0}\gtrsim100$, respectively. The former constraint translates into an upper-limit on the total baryon mass of $m_b \lesssim 1.8\times10^{14}$ g, which is many orders of magnitude smaller than that inferred from the afterglow observation of historical giant flare from SGR 1806--20: $m_b\sim10^{20}$--$10^{23}\ {\rm g}$ \citep{nakar05} or $m_b\gtrsim10^{24}\ {\rm g}$ \citep{granot06}.
A more precise time coincidence between the radio and the X-ray burst spikes (say $\Delta t_{\rm X,radio}\lesssim 1$ ms), if confirmed by a joint radio-X-ray timing analysis, would place stringent constraints on the baryon load and initial magnetization of the outflow. We defer the investigations of more realistic but complicated outflow models that include both baryons and magnetization within the GRB context (e.g., \citealt{gao15}) for future works.

Our results may have important implications for why the hard X-ray burst and coherent radio burst seen in April 27 event is rarely observed.
An interesting possibility is that magnetar flares launch relativistic outflows with different properties (e.g., degrees of baryon load and magnetization) and/or beaming \citep[e.g.,][]{fast_upperlimit_paper,zhang21}. 
In this case, the radiative efficiency changes from burst to burst. For example, one can speculate that the April 27 event might have loaded baryons. Correspondingly, the radial regions for fast-cooling can be expanded, enabling the hard X-ray emission. The diversity can be expected if ordinary flare events typically launch quasi-leptonic outflows (or even do not launch any outflow).

Finally, we encourage the search for the counterpart emissions at different wavelengths on different timescales. In the framework of ``burst-in-bubble'' model outlined by \cite{murase16},
relativistic outflows associated with the April 28 event may eventually collide with the nebula, leading to afterglow emission at multi-wavelengths. 
Future searches will be important for probing relativistic outflows with properties constrained by this work.

During finalizing the manuscript, we became aware of \citet{ioka20}, in which a formation of an extremely optically-thick trapped FB ($T_{\rm obs}=30$ keV) near the bottom of open magnetic field lines is considered. This special trapped FB powers an outflow that accelerates along the open magnetic field lines, which would generate the hard X-ray burst through diffusion of the X-mode FB photons. As discussed in \S \ref{ss:trappedFB}, such a scenario might be an interesting alternative to the possibility of generating hard X-ray bursts by multiple resonant scattering of original emission from an ordinary trapped FB ($T_{\rm obs}\sim10$ keV).

\section*{Acknowledgements}

We thank the anonymous referee for useful comments that improved the manuscript. We also thank Anna Ridnaia for sharing detailed information of the hard X-ray burst detected by Konus-Wind, Yuri Lyubarsky for a useful conversation on the evolution of magnetized outflow and Kunihito Ioka for useful comments on the effect of global magnetic field that can affect the early evolution of the outflow. The work of SY was supported by the Institute for Cosmic Ray Research and the advanced ERC grant TReX. The work of K.K. is supported by KAKENHI No.~20K04010. 
The work of K.M. is supported by the NSF Grant No.~AST-1908689, No.~AST-2108466 and No.~AST-2108467, and KAKENHI No.~20H01901 and No.~20H05852.

\section*{Data Availability}
This is a theoretical paper that does not involve any new data.

\bibliographystyle{mnras}

\begin{thebibliography}{}
\makeatletter
\relax
\def\mn@urlcharsother{\let\do\@makeother \do\$\do\&\do\#\do\^\do\_\do\%\do\~}
\def\mn@doi{\begingroup\mn@urlcharsother \@ifnextchar [ {\mn@doi@}
  {\mn@doi@[]}}
\def\mn@doi@[#1]#2{\def\@tempa{#1}\ifx\@tempa\@empty \href
  {http://dx.doi.org/#2} {doi:#2}\else \href {http://dx.doi.org/#2} {#1}\fi
  \endgroup}
\def\mn@eprint#1#2{\mn@eprint@#1:#2::\@nil}
\def\mn@eprint@arXiv#1{\href {http://arxiv.org/abs/#1} {{\tt arXiv:#1}}}
\def\mn@eprint@dblp#1{\href {http://dblp.uni-trier.de/rec/bibtex/#1.xml}
  {dblp:#1}}
\def\mn@eprint@#1:#2:#3:#4\@nil{\def\@tempa {#1}\def\@tempb {#2}\def\@tempc
  {#3}\ifx \@tempc \@empty \let \@tempc \@tempb \let \@tempb \@tempa \fi \ifx
  \@tempb \@empty \def\@tempb {arXiv}\fi \@ifundefined
  {mn@eprint@\@tempb}{\@tempb:\@tempc}{\expandafter \expandafter \csname
  mn@eprint@\@tempb\endcsname \expandafter{\@tempc}}}

\bibitem[\protect\citeauthoryear{{Arons} \& {Barnard}}{{Arons} \&
  {Barnard}}{1986}]{arons86}
{Arons} J.,  {Barnard} J.~J.,  1986, \mn@doi [\apj] {10.1086/163978}, \href
  {https://ui.adsabs.harvard.edu/abs/1986ApJ...302..120A} {302, 120}

\bibitem[\protect\citeauthoryear{{Beloborodov}}{{Beloborodov}}{2013}]{beloborodov13}
{Beloborodov} A.~M.,  2013, \mn@doi [\apj] {10.1088/0004-637X/777/2/114}, \href
  {https://ui.adsabs.harvard.edu/abs/2013ApJ...777..114B} {777, 114}

\bibitem[\protect\citeauthoryear{{Beloborodov}}{{Beloborodov}}{2017}]{beloborodov17}
{Beloborodov} A.~M.,  2017, \mn@doi [\apjl] {10.3847/2041-8213/aa78f3}, \href
  {https://ui.adsabs.harvard.edu/abs/2017ApJ...843L..26B} {843, L26}

\bibitem[\protect\citeauthoryear{{Beloborodov}}{{Beloborodov}}{2020}]{beloborodov20}
{Beloborodov} A.~M.,  2020, \mn@doi [\apj] {10.3847/1538-4357/ab83eb}, \href
  {https://ui.adsabs.harvard.edu/abs/2020ApJ...896..142B} {896, 142}

\bibitem[\protect\citeauthoryear{{Beloborodov} \& {Thompson}}{{Beloborodov} \&
  {Thompson}}{2007}]{beloborodov07}
{Beloborodov} A.~M.,  {Thompson} C.,  2007, \mn@doi [\apj] {10.1086/508917},
  \href {https://ui.adsabs.harvard.edu/abs/2007ApJ...657..967B} {657, 967}

\bibitem[\protect\citeauthoryear{{Beskin}, {Kuznetsova}  \& {Rafikov}}{{Beskin}
  et~al.}{1998}]{beskin98}
{Beskin} V.~S.,  {Kuznetsova} I.~V.,   {Rafikov} R.~R.,  1998, \mn@doi [\mnras]
  {10.1046/j.1365-8711.1998.01659.x}, \href
  {https://ui.adsabs.harvard.edu/abs/1998MNRAS.299..341B} {299, 341}

\bibitem[\protect\citeauthoryear{{Bochenek}, {Ravi}, {Belov}, {Hallinan},
  {Kocz}, {Kulkarni}  \& {McKenna}}{{Bochenek} et~al.}{2020}]{stare2_paper}
{Bochenek} C.~D.,  {Ravi} V.,  {Belov} K.~V.,  {Hallinan} G.,  {Kocz} J.,
  {Kulkarni} S.~R.,   {McKenna} D.~L.,  2020, arXiv e-prints, \href
  {https://ui.adsabs.harvard.edu/abs/2020arXiv200510828B} {p. arXiv:2005.10828}

\bibitem[\protect\citeauthoryear{{Borghese}, {Coti Zelati}, {Rea}, {Esposito},
  {Israel}, {Mereghetti}  \& {Tiengo}}{{Borghese} et~al.}{2020}]{borghese20}
{Borghese} A.,  {Coti Zelati} F.,  {Rea} N.,  {Esposito} P.,  {Israel} G.~L.,
  {Mereghetti} S.,   {Tiengo} A.,  2020, \mn@doi [\apjl]
  {10.3847/2041-8213/aba82a}, \href
  {https://ui.adsabs.harvard.edu/abs/2020ApJ...902L...2B} {902, L2}

\bibitem[\protect\citeauthoryear{{CHIME/FRB Collaboration} et~al.,}{{CHIME/FRB
  Collaboration} et~al.}{2020}]{chime_paper}
{CHIME/FRB Collaboration} et~al., 2020, \mn@doi [\nat]
  {10.1038/s41586-020-2863-y}, \href
  {https://ui.adsabs.harvard.edu/abs/2020Natur.587...54C} {587, 54}

\bibitem[\protect\citeauthoryear{{Canuto} \& {Ventura}}{{Canuto} \&
  {Ventura}}{1977}]{canuto77}
{Canuto} V.,  {Ventura} J.,  1977, \fcp, \href
  {https://ui.adsabs.harvard.edu/abs/1977FCPh....2..203C} {2, 203}

\bibitem[\protect\citeauthoryear{Chen}{Chen}{1984}]{chen84}
Chen F.~F.,  1984, Introduction to Plasma Physics and controlled fusion, 2nd
  edition (Plenum Press, New York)

\bibitem[\protect\citeauthoryear{{Contopoulos} \& {Kazanas}}{{Contopoulos} \&
  {Kazanas}}{2002}]{contopoulos02}
{Contopoulos} I.,  {Kazanas} D.,  2002, \mn@doi [\apj] {10.1086/324778}, \href
  {https://ui.adsabs.harvard.edu/abs/2002ApJ...566..336C} {566, 336}

\bibitem[\protect\citeauthoryear{{DeLaunay} et~al.,}{{DeLaunay}
  et~al.}{2016}]{2016ApJ...832L...1D}
{DeLaunay} J.~J.,  et~al., 2016, \mn@doi [\apjl] {10.3847/2041-8205/832/1/L1},
  \href {https://ui.adsabs.harvard.edu/abs/2016ApJ...832L...1D} {832, L1}

\bibitem[\protect\citeauthoryear{{Drenkhahn}}{{Drenkhahn}}{2002}]{drenkhahn02}
{Drenkhahn} G.,  2002, \mn@doi [\aap] {10.1051/0004-6361:20020390}, \href
  {https://ui.adsabs.harvard.edu/abs/2002A&A...387..714D} {387, 714}

\bibitem[\protect\citeauthoryear{{Drenkhahn} \& {Spruit}}{{Drenkhahn} \&
  {Spruit}}{2002}]{ds02}
{Drenkhahn} G.,  {Spruit} H.~C.,  2002, \mn@doi [\aap]
  {10.1051/0004-6361:20020839}, \href
  {https://ui.adsabs.harvard.edu/abs/2002A&A...391.1141D} {391, 1141}

\bibitem[\protect\citeauthoryear{{Eichler} \& {Waxman}}{{Eichler} \&
  {Waxman}}{2005}]{eichler05}
{Eichler} D.,  {Waxman} E.,  2005, \mn@doi [\apj] {10.1086/430596}, \href
  {https://ui.adsabs.harvard.edu/abs/2005ApJ...627..861E} {627, 861}

\bibitem[\protect\citeauthoryear{{Gao} \& {Zhang}}{{Gao} \&
  {Zhang}}{2015}]{gao15}
{Gao} H.,  {Zhang} B.,  2015, \mn@doi [\apj] {10.1088/0004-637X/801/2/103},
  \href {https://ui.adsabs.harvard.edu/abs/2015ApJ...801..103G} {801, 103}

\bibitem[\protect\citeauthoryear{{Ghisellini} \& {Locatelli}}{{Ghisellini} \&
  {Locatelli}}{2018}]{ghisellini18}
{Ghisellini} G.,  {Locatelli} N.,  2018, \mn@doi [\aap]
  {10.1051/0004-6361/201731820}, \href
  {https://ui.adsabs.harvard.edu/abs/2018A&A...613A..61G} {613, A61}

\bibitem[\protect\citeauthoryear{{Goodman}}{{Goodman}}{1986}]{goodman86}
{Goodman} J.,  1986, \mn@doi [\apjl] {10.1086/184741}, \href
  {http://adsabs.harvard.edu/abs/1986ApJ...308L..47G} {308, L47}

\bibitem[\protect\citeauthoryear{{Granot} et~al.,}{{Granot}
  et~al.}{2006}]{granot06}
{Granot} J.,  et~al., 2006, \mn@doi [\apj] {10.1086/497680}, \href
  {http://adsabs.harvard.edu/abs/2006ApJ...638..391G} {638, 391}

\bibitem[\protect\citeauthoryear{{Grimsrud} \& {Wasserman}}{{Grimsrud} \&
  {Wasserman}}{1998}]{grimsrud98}
{Grimsrud} O.~M.,  {Wasserman} I.,  1998, \mn@doi [\mnras]
  {10.1046/j.1365-8711.1998.02004.x}, \href
  {http://adsabs.harvard.edu/abs/1998MNRAS.300.1158G} {300, 1158}

\bibitem[\protect\citeauthoryear{{Ioka}}{{Ioka}}{2020}]{ioka20}
{Ioka} K.,  2020, \mn@doi [\apjl] {10.3847/2041-8213/abc6a3}, \href
  {https://ui.adsabs.harvard.edu/abs/2020ApJ...904L..15I} {904, L15}

\bibitem[\protect\citeauthoryear{{Israel} et~al.,}{{Israel}
  et~al.}{2016}]{israel16}
{Israel} G.~L.,  et~al., 2016, \mn@doi [\mnras] {10.1093/mnras/stw008}, \href
  {https://ui.adsabs.harvard.edu/abs/2016MNRAS.457.3448I} {457, 3448}

\bibitem[\protect\citeauthoryear{{Iwamoto}, {Amano}, {Hoshino}  \&
  {Matsumoto}}{{Iwamoto} et~al.}{2017}]{iwamoto17}
{Iwamoto} M.,  {Amano} T.,  {Hoshino} M.,   {Matsumoto} Y.,  2017, \mn@doi
  [\apj] {10.3847/1538-4357/aa6d6f}, \href
  {https://ui.adsabs.harvard.edu/abs/2017ApJ...840...52I} {840, 52}

\bibitem[\protect\citeauthoryear{{Iwamoto}, {Amano}, {Hoshino}, {Matsumoto},
  {Niemiec}, {Ligorini}, {Kobzar}  \& {Pohl}}{{Iwamoto}
  et~al.}{2019}]{iwamoto19}
{Iwamoto} M.,  {Amano} T.,  {Hoshino} M.,  {Matsumoto} Y.,  {Niemiec} J.,
  {Ligorini} A.,  {Kobzar} O.,   {Pohl} M.,  2019, \mn@doi [\apjl]
  {10.3847/2041-8213/ab4265}, \href
  {https://ui.adsabs.harvard.edu/abs/2019ApJ...883L..35I} {883, L35}

\bibitem[\protect\citeauthoryear{{Katz}}{{Katz}}{2016}]{katz16}
{Katz} J.~I.,  2016, \mn@doi [\apj] {10.3847/0004-637X/826/2/226}, \href
  {https://ui.adsabs.harvard.edu/abs/2016ApJ...826..226K} {826, 226}

\bibitem[\protect\citeauthoryear{{Katz}}{{Katz}}{2020}]{katz20}
{Katz} J.~I.,  2020, \mn@doi [\mnras] {10.1093/mnras/staa3042}, \href
  {https://ui.adsabs.harvard.edu/abs/2020MNRAS.499.2319K} {499, 2319}

\bibitem[\protect\citeauthoryear{{Kennel} \& {Coroniti}}{{Kennel} \&
  {Coroniti}}{1984}]{kc84}
{Kennel} C.~F.,  {Coroniti} F.~V.,  1984, \mn@doi [\apj] {10.1086/162357},
  \href {https://ui.adsabs.harvard.edu/abs/1984ApJ...283..710K} {283, 710}

\bibitem[\protect\citeauthoryear{{Kirsten}, {Snelders}, {Jenkins}, {Nimmo},
  {van den Eijnden}, {Hessels}, {Gawro{\'n}ski}  \& {Yang}}{{Kirsten}
  et~al.}{2021}]{kirsten20}
{Kirsten} F.,  {Snelders} M.~P.,  {Jenkins} M.,  {Nimmo} K.,  {van den Eijnden}
  J.,  {Hessels} J.~W.~T.,  {Gawro{\'n}ski} M.~P.,   {Yang} J.,  2021, \mn@doi
  [Nature Astronomy] {10.1038/s41550-020-01246-3}, \href
  {https://ui.adsabs.harvard.edu/abs/2021NatAs...5..414K} {5, 414}

\bibitem[\protect\citeauthoryear{{Komissarov}, {Vlahakis}, {K{\"o}nigl}  \&
  {Barkov}}{{Komissarov} et~al.}{2009}]{komissarov09}
{Komissarov} S.~S.,  {Vlahakis} N.,  {K{\"o}nigl} A.,   {Barkov} M.~V.,  2009,
  \mn@doi [\mnras] {10.1111/j.1365-2966.2009.14410.x}, \href
  {https://ui.adsabs.harvard.edu/abs/2009MNRAS.394.1182K} {394, 1182}

\bibitem[\protect\citeauthoryear{{Kothes}, {Sun}, {Gaensler}  \&
  {Reich}}{{Kothes} et~al.}{2018}]{kothes18}
{Kothes} R.,  {Sun} X.,  {Gaensler} B.,   {Reich} W.,  2018, \mn@doi [\apj]
  {10.3847/1538-4357/aa9e89}, \href
  {https://ui.adsabs.harvard.edu/abs/2018ApJ...852...54K} {852, 54}

\bibitem[\protect\citeauthoryear{{Kumar}, {Lu}  \& {Bhattacharya}}{{Kumar}
  et~al.}{2017}]{kumar17}
{Kumar} P.,  {Lu} W.,   {Bhattacharya} M.,  2017, \mn@doi [\mnras]
  {10.1093/mnras/stx665}, \href
  {https://ui.adsabs.harvard.edu/abs/2017MNRAS.468.2726K} {468, 2726}

\bibitem[\protect\citeauthoryear{{Li} et~al.,}{{Li} et~al.}{2021a}]{hxmt_paper}
{Li} C.~K.,  et~al., 2021a, Nature Astronomy, \href
  {https://ui.adsabs.harvard.edu/abs/2021NatAs...5..378L} {5, 378}

\bibitem[\protect\citeauthoryear{Li et~al.}{Li et~al.}{2021b}]{li21}
Li D.,  et~al., 2021b, \mn@doi [Nature] {10.1038/s41586-021-03878-5}, 598, 267

\bibitem[\protect\citeauthoryear{{Lin} et~al.,}{{Lin}
  et~al.}{2020}]{fast_upperlimit_paper}
{Lin} L.,  et~al., 2020, \mn@doi [\nat] {10.1038/s41586-020-2839-y}, \href
  {https://ui.adsabs.harvard.edu/abs/2020Natur.587...63L} {587, 63}

\bibitem[\protect\citeauthoryear{{Lithwick} \& {Sari}}{{Lithwick} \&
  {Sari}}{2001}]{lithwick01}
{Lithwick} Y.,  {Sari} R.,  2001, \mn@doi [\apj] {10.1086/321455}, \href
  {https://ui.adsabs.harvard.edu/abs/2001ApJ...555..540L} {555, 540}

\bibitem[\protect\citeauthoryear{{Lu} \& {Kumar}}{{Lu} \& {Kumar}}{2018}]{lu18}
{Lu} W.,  {Kumar} P.,  2018, \mn@doi [\mnras] {10.1093/mnras/sty716}, \href
  {https://ui.adsabs.harvard.edu/abs/2018MNRAS.477.2470L} {477, 2470}

\bibitem[\protect\citeauthoryear{{Lu}, {Kumar}  \& {Zhang}}{{Lu}
  et~al.}{2020}]{lu20}
{Lu} W.,  {Kumar} P.,   {Zhang} B.,  2020, \mn@doi [\mnras]
  {10.1093/mnras/staa2450}, \href
  {https://ui.adsabs.harvard.edu/abs/2020MNRAS.498.1397L} {498, 1397}

\bibitem[\protect\citeauthoryear{Luo et~al.}{Luo et~al.}{2020}]{luo20}
Luo R.,  et~al., 2020, \mn@doi [Nature] {10.1038/s41586-020-2827-2}, 586, 693

\bibitem[\protect\citeauthoryear{{Lyubarsky}}{{Lyubarsky}}{2002}]{lyubarsky02}
{Lyubarsky} Y.~E.,  2002, \mn@doi [\mnras] {10.1046/j.1365-8711.2002.05290.x},
  \href {https://ui.adsabs.harvard.edu/abs/2002MNRAS.332..199L} {332, 199}

\bibitem[\protect\citeauthoryear{{Lyubarsky}}{{Lyubarsky}}{2014}]{lyubarsky14}
{Lyubarsky} Y.,  2014, \mn@doi [\mnras] {10.1093/mnrasl/slu046}, \href
  {https://ui.adsabs.harvard.edu/abs/2014MNRAS.442L...9L} {442, L9}

\bibitem[\protect\citeauthoryear{Lyubarsky}{Lyubarsky}{2021}]{lyubarski21}
Lyubarsky Y.,  2021, \mn@doi [Universe] {10.3390/universe7030056}, 7

\bibitem[\protect\citeauthoryear{{Lyubarsky} \& {Kirk}}{{Lyubarsky} \&
  {Kirk}}{2001}]{lyubarsky01}
{Lyubarsky} Y.,  {Kirk} J.~G.,  2001, \mn@doi [\apj] {10.1086/318354}, \href
  {https://ui.adsabs.harvard.edu/abs/2001ApJ...547..437L} {547, 437}

\bibitem[\protect\citeauthoryear{{Lyutikov}}{{Lyutikov}}{2020}]{lyutikov20b}
{Lyutikov} M.,  2020, arXiv e-prints, \href
  {https://ui.adsabs.harvard.edu/abs/2020arXiv200616029L} {p. arXiv:2006.16029}

\bibitem[\protect\citeauthoryear{{Lyutikov} \& {Popov}}{{Lyutikov} \&
  {Popov}}{2020}]{lyutikov20}
{Lyutikov} M.,  {Popov} S.,  2020, arXiv e-prints, \href
  {https://ui.adsabs.harvard.edu/abs/2020arXiv200505093L} {p. arXiv:2005.05093}

\bibitem[\protect\citeauthoryear{{Margalit}, {Beniamini}, {Sridhar}  \&
  {Metzger}}{{Margalit} et~al.}{2020}]{margalit20}
{Margalit} B.,  {Beniamini} P.,  {Sridhar} N.,   {Metzger} B.~D.,  2020,
  \mn@doi [\apjl] {10.3847/2041-8213/abac57}, \href
  {https://ui.adsabs.harvard.edu/abs/2020ApJ...899L..27M} {899, L27}

\bibitem[\protect\citeauthoryear{{Matsumoto}, {Nakar}  \& {Piran}}{{Matsumoto}
  et~al.}{2019}]{matsumoto19}
{Matsumoto} T.,  {Nakar} E.,   {Piran} T.,  2019, \mn@doi [\mnras]
  {10.1093/mnras/stz923}, \href
  {https://ui.adsabs.harvard.edu/abs/2019MNRAS.486.1563M} {486, 1563}

\bibitem[\protect\citeauthoryear{{Mereghetti} et~al.,}{{Mereghetti}
  et~al.}{2020}]{integral_paper}
{Mereghetti} S.,  et~al., 2020, \mn@doi [\apjl] {10.3847/2041-8213/aba2cf},
  \href {https://ui.adsabs.harvard.edu/abs/2020ApJ...898L..29M} {898, L29}

\bibitem[\protect\citeauthoryear{{Meszaros}}{{Meszaros}}{1992}]{meszaros92}
{Meszaros} P.,  1992, {High-energy radiation from magnetized neutron stars}

\bibitem[\protect\citeauthoryear{{Metzger}, {Margalit}  \& {Sironi}}{{Metzger}
  et~al.}{2019}]{metzger19}
{Metzger} B.~D.,  {Margalit} B.,   {Sironi} L.,  2019, \mn@doi [\mnras]
  {10.1093/mnras/stz700}, \href
  {https://ui.adsabs.harvard.edu/abs/2019MNRAS.485.4091M} {485, 4091}

\bibitem[\protect\citeauthoryear{{Murase}, {Kashiyama}  \&
  {M{\'e}sz{\'a}ros}}{{Murase} et~al.}{2016}]{murase16}
{Murase} K.,  {Kashiyama} K.,   {M{\'e}sz{\'a}ros} P.,  2016, \mn@doi [\mnras]
  {10.1093/mnras/stw1328}, \href
  {http://adsabs.harvard.edu/abs/2016MNRAS.461.1498M} {461, 1498}

\bibitem[\protect\citeauthoryear{{Murase}, {M{\'e}sz{\'a}ros}  \&
  {Fox}}{{Murase} et~al.}{2017}]{murase17}
{Murase} K.,  {M{\'e}sz{\'a}ros} P.,   {Fox} D.~B.,  2017, \mn@doi [\apjl]
  {10.3847/2041-8213/836/1/L6}, \href
  {https://ui.adsabs.harvard.edu/abs/2017ApJ...836L...6M} {836, L6}

\bibitem[\protect\citeauthoryear{{Nakar}}{{Nakar}}{2007}]{nakar07}
{Nakar} E.,  2007, \mn@doi [\physrep] {10.1016/j.physrep.2007.02.005}, \href
  {https://ui.adsabs.harvard.edu/abs/2007PhR...442..166N} {442, 166}

\bibitem[\protect\citeauthoryear{{Nakar}, {Piran}  \& {Sari}}{{Nakar}
  et~al.}{2005}]{nakar05}
{Nakar} E.,  {Piran} T.,   {Sari} R.,  2005, \mn@doi [\apj] {10.1086/497296},
  \href {https://ui.adsabs.harvard.edu/abs/2005ApJ...635..516N} {635, 516}

\bibitem[\protect\citeauthoryear{{Paczynski}}{{Paczynski}}{1986}]{paczynski86}
{Paczynski} B.,  1986, \mn@doi [\apjl] {10.1086/184740}, \href
  {http://adsabs.harvard.edu/abs/1986ApJ...308L..43P} {308, L43}

\bibitem[\protect\citeauthoryear{{Piran}}{{Piran}}{1999}]{piran99}
{Piran} T.,  1999, \mn@doi [\physrep] {10.1016/S0370-1573(98)00127-6}, \href
  {https://ui.adsabs.harvard.edu/abs/1999PhR...314..575P} {314, 575}

\bibitem[\protect\citeauthoryear{{Plotnikov} \& {Sironi}}{{Plotnikov} \&
  {Sironi}}{2019}]{plotnikov19}
{Plotnikov} I.,  {Sironi} L.,  2019, \mn@doi [\mnras] {10.1093/mnras/stz640},
  \href {https://ui.adsabs.harvard.edu/abs/2019MNRAS.485.3816P} {485, 3816}

\bibitem[\protect\citeauthoryear{{Ridnaia} et~al.,}{{Ridnaia}
  et~al.}{2021}]{konus_paper}
{Ridnaia} A.,  et~al., 2021, \mn@doi [Nature Astronomy]
  {10.1038/s41550-020-01265-0}, \href
  {https://ui.adsabs.harvard.edu/abs/2021NatAs...5..372R} {5, 372}

\bibitem[\protect\citeauthoryear{{Sari}, {Piran}  \& {Narayan}}{{Sari}
  et~al.}{1998}]{sari98}
{Sari} R.,  {Piran} T.,   {Narayan} R.,  1998, \mn@doi [\apjl]
  {10.1086/311269}, \href
  {https://ui.adsabs.harvard.edu/abs/1998ApJ...497L..17S} {497, L17}

\bibitem[\protect\citeauthoryear{{Shemi} \& {Piran}}{{Shemi} \&
  {Piran}}{1990}]{shemi90}
{Shemi} A.,  {Piran} T.,  1990, \mn@doi [\apjl] {10.1086/185887}, \href
  {https://ui.adsabs.harvard.edu/abs/1990ApJ...365L..55S} {365, L55}

\bibitem[\protect\citeauthoryear{{Simard} \& {Ravi}}{{Simard} \&
  {Ravi}}{2020}]{simard20}
{Simard} D.,  {Ravi} V.,  2020, \mn@doi [\apjl] {10.3847/2041-8213/abaa40},
  \href {https://ui.adsabs.harvard.edu/abs/2020ApJ...899L..21S} {899, L21}

\bibitem[\protect\citeauthoryear{{Tavani} et~al.,}{{Tavani}
  et~al.}{2021}]{agile_paper}
{Tavani} M.,  et~al., 2021, \mn@doi [Nature Astronomy]
  {10.1038/s41550-020-01276-x}, \href
  {https://ui.adsabs.harvard.edu/abs/2021NatAs...5..401T} {5, 401}

\bibitem[\protect\citeauthoryear{{Thompson} \& {Duncan}}{{Thompson} \&
  {Duncan}}{1995}]{td95}
{Thompson} C.,  {Duncan} R.~C.,  1995, \mn@doi [\mnras]
  {10.1093/mnras/275.2.255}, \href
  {https://ui.adsabs.harvard.edu/abs/1995MNRAS.275..255T} {275, 255}

\bibitem[\protect\citeauthoryear{{Thompson} \& {Duncan}}{{Thompson} \&
  {Duncan}}{1996}]{td96}
{Thompson} C.,  {Duncan} R.~C.,  1996, \mn@doi [\apj] {10.1086/178147}, \href
  {http://adsabs.harvard.edu/abs/1996ApJ...473..322T} {473, 322}

\bibitem[\protect\citeauthoryear{{Thompson}, {Lyutikov}  \&
  {Kulkarni}}{{Thompson} et~al.}{2002}]{tlk02}
{Thompson} C.,  {Lyutikov} M.,   {Kulkarni} S.~R.,  2002, \mn@doi [\apj]
  {10.1086/340586}, \href
  {https://ui.adsabs.harvard.edu/abs/2002ApJ...574..332T} {574, 332}

\bibitem[\protect\citeauthoryear{{Uzdensky}, {Loureiro}  \&
  {Schekochihin}}{{Uzdensky} et~al.}{2010}]{uzdensky10}
{Uzdensky} D.~A.,  {Loureiro} N.~F.,   {Schekochihin} A.~A.,  2010, \mn@doi
  [\prl] {10.1103/PhysRevLett.105.235002}, \href
  {https://ui.adsabs.harvard.edu/abs/2010PhRvL.105w5002U} {105, 235002}

\bibitem[\protect\citeauthoryear{{Wadiasingh} \& {Chirenti}}{{Wadiasingh} \&
  {Chirenti}}{2020}]{wadiasingh20}
{Wadiasingh} Z.,  {Chirenti} C.,  2020, \mn@doi [\apjl]
  {10.3847/2041-8213/abc562}, \href
  {https://ui.adsabs.harvard.edu/abs/2020ApJ...903L..38W} {903, L38}

\bibitem[\protect\citeauthoryear{{Wadiasingh} \& {Timokhin}}{{Wadiasingh} \&
  {Timokhin}}{2019}]{wadiasingh19}
{Wadiasingh} Z.,  {Timokhin} A.,  2019, \mn@doi [\apj]
  {10.3847/1538-4357/ab2240}, \href
  {https://ui.adsabs.harvard.edu/abs/2019ApJ...879....4W} {879, 4}

\bibitem[\protect\citeauthoryear{{Wang}}{{Wang}}{2020}]{wang20}
{Wang} J.-S.,  2020, \mn@doi [\apj] {10.3847/1538-4357/aba955}, \href
  {https://ui.adsabs.harvard.edu/abs/2020ApJ...900..172W} {900, 172}

\bibitem[\protect\citeauthoryear{{Waxman}}{{Waxman}}{2017}]{waxman2017}
{Waxman} E.,  2017, \mn@doi [\apj] {10.3847/1538-4357/aa713e}, \href
  {https://ui.adsabs.harvard.edu/abs/2017ApJ...842...34W} {842, 34}

\bibitem[\protect\citeauthoryear{{Yamasaki}, {Kisaka}, {Terasawa}  \&
  {Enoto}}{{Yamasaki} et~al.}{2019}]{yamasaki19}
{Yamasaki} S.,  {Kisaka} S.,  {Terasawa} T.,   {Enoto} T.,  2019, \mn@doi
  [\mnras] {10.1093/mnras/sty3388}, \href
  {https://ui.adsabs.harvard.edu/abs/2019MNRAS.483.4175Y} {483, 4175}

\bibitem[\protect\citeauthoryear{{Yamasaki}, {Lyubarsky}, {Granot}  \&
  {G{\"o}{\u{g}}{\"u}{\c{s}}}}{{Yamasaki} et~al.}{2020}]{yamasaki20}
{Yamasaki} S.,  {Lyubarsky} Y.,  {Granot} J.,   {G{\"o}{\u{g}}{\"u}{\c{s}}} E.,
   2020, \mn@doi [\mnras] {10.1093/mnras/staa2223}, \href
  {https://ui.adsabs.harvard.edu/abs/2020MNRAS.498..484Y} {498, 484}

\bibitem[\protect\citeauthoryear{{Yang} \& {Zhang}}{{Yang} \&
  {Zhang}}{2018}]{yang18}
{Yang} Y.-P.,  {Zhang} B.,  2018, \mn@doi [\apj] {10.3847/1538-4357/aae685},
  \href {https://ui.adsabs.harvard.edu/abs/2018ApJ...868...31Y} {868, 31}

\bibitem[\protect\citeauthoryear{{Yang} \& {Zhang}}{{Yang} \&
  {Zhang}}{2021}]{yang21}
{Yang} Y.-P.,  {Zhang} B.,  2021, arXiv e-prints, \href
  {https://ui.adsabs.harvard.edu/abs/2021arXiv210401925Y} {p. arXiv:2104.01925}

\bibitem[\protect\citeauthoryear{{Younes} et~al.,}{{Younes}
  et~al.}{2021}]{nicergbm_paper}
{Younes} G.,  et~al., 2021, Nature Astronomy, \href
  {https://ui.adsabs.harvard.edu/abs/2021NatAs...5..408Y} {5, 408}

\bibitem[\protect\citeauthoryear{{Yu}, {Zou}, {Dai}  \& {Yu}}{{Yu}
  et~al.}{2021}]{yu20}
{Yu} Y.-W.,  {Zou} Y.-C.,  {Dai} Z.-G.,   {Yu} W.-F.,  2021, \mn@doi [\mnras]
  {10.1093/mnras/staa3374}, \href
  {https://ui.adsabs.harvard.edu/abs/2021MNRAS.500.2704Y} {500, 2704}

\bibitem[\protect\citeauthoryear{{Yuan}, {Beloborodov}, {Chen}  \&
  {Levin}}{{Yuan} et~al.}{2020}]{yuan20}
{Yuan} Y.,  {Beloborodov} A.~M.,  {Chen} A.~Y.,   {Levin} Y.,  2020, \mn@doi
  [\apjl] {10.3847/2041-8213/abafa8}, \href
  {https://ui.adsabs.harvard.edu/abs/2020ApJ...900L..21Y} {900, L21}

\bibitem[\protect\citeauthoryear{{Zhang}}{{Zhang}}{2021}]{zhang21}
{Zhang} B.,  2021, \mn@doi [\apjl] {10.3847/2041-8213/abd628}, \href
  {https://ui.adsabs.harvard.edu/abs/2021ApJ...907L..17Z} {907, L17}

\bibitem[\protect\citeauthoryear{{Zhou}, {Zhou}, {Chen}, {Wang}, {Vink}  \&
  {Wang}}{{Zhou} et~al.}{2020}]{zhou20}
{Zhou} P.,  {Zhou} X.,  {Chen} Y.,  {Wang} J.-S.,  {Vink} J.,   {Wang} Y.,
  2020, \mn@doi [\apj] {10.3847/1538-4357/abc34a}, \href
  {https://ui.adsabs.harvard.edu/abs/2020ApJ...905...99Z} {905, 99}

\makeatother
\end{thebibliography}
\input{GalacticFRB.bbl}

\appendix

\section{Relativistic Outflow Models}
\label{s:appendix}
\subsection{Leptonic Wind}
\label{ss:leptonic wind}
First let us consider an outflow composed of $e^{\pm}$ pairs plus photons. In order to track the evolution of pure-leptonic FB, we follow the formulation by \citet{grimsrud98} who considered non-equilibrium effects that would modify the early pair density evolution (see also Appendix of \citealt{yamasaki19}). 
The conservation of energy, momentum and pair number density for a steady flow in spherical symmetry leads to a set of simple scaling laws that govern the radial evolution of the bulk Lorentz factor and temperature
(\citealt{paczynski86,goodman86,shemi90}). The bulk Lorentz factor increases linearly with $r$ as
$\Gamma\approx\Gamma_0(r/r_0)$ for $r<r_{\infty}$, where $r_0$ is the initial FB size and $r_{\infty}$ the saturation radius above which the acceleration of plasma stops and FB enters a coasting phase with an asymptotic bulk Lorentz factor $\Gamma_{\infty}$. Meanwhile the FB temperature cools as $T^{\prime}\approx T_0(r/r_0)^{-1}$. The dynamical evolution of FB is uniquely determined by initial conditions, i.e., a size $r_0$, temperature $T_0$ and Lorentz factor $\Gamma_0$. We relate the initial parameters to the total outflow energy $E_{\rm flare}$ by
\beqn
\label{eq:E_fb}
E_{\rm flare}=\Gamma_0 aT_0^4r_0^3\sim10^{40}\,r_{0,5}^3\, \Theta_0^4 \ {\rm erg},
\eeqn
where we adopt reference values as $r_0=R_0\sim10^5$ cm and $E_{\rm flare}\sim 10^{40} {\ \rm erg}$ based on the trapped FB parameters estimated in \S \ref{ss:trappedFB}.
In the second equality we  implicitly assume that the initial FB is at rest ($\Gamma_0=1$). Note that $\Theta_0\equiv T_0/m_ec^2$ denotes the dimensionless initial FB (outflow) temperature, which is set to be unity (rather than $\hat{\Theta}_0=0.4$ assumed in this work) here for purposes demonstration.

In addition to the dynamical evolution, we consider the evolution of the pair number density, taking into account the interactions among pairs and photons (i.e., creation and annihilation).
In the stage of expansion the FB plasma evolves with the non-magnetic equilibrium number density  
\beqn
\label{eq:n_eq}
n_{e,\rm eq}(T)\approx\frac{1}{\sqrt{2\pi^3}}\,\lambda_{\rm C}^{-3}\,\left(\frac{T}{m_ec^2}\right)^{3/2}\,e^{-m_ec^2/T}.
\eeqn
Compared to Eq. \eqref{eq:n_eq_mag}, the magnetic term vanishes and the temperature dependence changes. Starting from $n_{e,0}^{\prime}=n_{e,\rm eq}(T_0)$, the radial evolution of electron (positron) number density is summarized below.

The initial FB is at rest in pair equilibrium due to its high temperature with its size $r=r_0$. It immediately expands and cools down to the electron rest mass energy, and then $n_e$ begins to deviate from $n_{e,\rm eq}$. The pair annihilation dominates the pair process since the number of pair-creating high-energy photons decreases as the FB cools. Eventually, the FB reaches the photospheric radius $r_{\rm ph}\sim2.5\times10^6\ {\rm cm}\ r_{0,5}\,\Theta_0$ at which the optical depth to electron scattering becomes an order of unity.
When the FB becomes optically thin, photons begin to leak freely out of the photosphere. However, they still continues to supply the radiation energy to pairs, which accelerates pairs up to the coasting radius $r_{\infty}\sim1.1\times10^8\ {\rm cm}\ r_{0,5}^{5/4}\, \Theta_0$. The photons cease to inject the radiation energy to pairs, and the FB begins to freely coast at constant speed $\Gamma_{\infty}=r_{\infty}/r_i\sim1.1\times10^3\ r_{0,5}^{1/4} \,\Theta_0$. At this stage, the pair annihilation no longer occurs due to the small number density. As a result, the total number of pairs conserves and the pair density evolves as $\propto r^{-2}$. The number density of the pair at the coasting phase has an analytical form \citep{yamasaki19}:
\beqn
\label{eq:ne_leptonic}
n_e^{\prime}(r)=5.5\times10^{30}\ r_{0,5}^{3/4}\,\Theta_0^2\, r^{-2} \ \ {\rm cm^{-3}},
\eeqn
which is valid for $r>r_{\infty}$. 
Consequently, the plasma cutoff radius $r_{\rm cutoff}$ at which $\nu_{\rm p}^{\rm maser}=\nu_{\rm obs}$ is
\beqn
\label{eq:nu_pl_leptonic}
r_{\rm cutoff}\sim2.3\times10^{13}\ r_{0,5}^{5/8}\,\Theta_0^2\, \zeta\,\nu_{\rm obs,9}^{-1} \ \ {\rm cm}.
\eeqn
Figure \ref{fig:leptonicFB} shows the overall evolution of leptonic FB.

\subsection{Baryonic Wind}
Provided that the FB outflow forms in the vicinity of the NS surface, it is expected that some amount of baryons might be contaminated, which was most likely the case for SGR 1806--20 giant flare in 2004 \citep{granot06}. This might affect the radial evolution of FB with respect to the pure-leptonic case (e.g., \citealt{grimsrud98,nakar05}). Conservation of baryon number and energy reads
\beqn
\label{eq: M_dot and E_dot}
    \dot{M}&=&r^2\rho^{\prime}\,\Gamma \beta c = const,\\
    L&=&r^2(U^{\prime}+P^{\prime})\Gamma^2\beta c=const,
\eeqn
where $\rho^{\prime}$, $U^{\prime}$ and $P^{\prime}$ are the rest mass density, the total energy density and the total pressure, respectively. In case of baryonic wind, $\rho^{\prime}=Am_pn^{\prime}$, where
$n^{\prime}$ is the comoving baryon number density with mass number $A$ (and  atomic number $Z$) and $m_p$ being the proton mass.
The magnitude of bulk Lorentz factor is limited by the total entropy per baryon in the FB as
\beqn
\label{eq:eta}
\eta\equiv\frac{L}{\dot{M}c^2}=\frac{(U^{\prime}+P^{\prime})\Gamma}{Am_pc^2 n^{\prime}}.
\eeqn
We can see that the adiabatic evolution ($\Gamma\propto r$ and $T^{\prime}\propto 1/r$) breaks up when the kinetic energy begins to dominate the radiation energy. This transition takes place when $U^{\prime}+P^{\prime}\sim Am_pn^{\prime}c^2$ with a corresponding radius $r_{\rm M}=\eta \,(r_0/\Gamma_0)$, above which the Lorentz factor stays constant ($\Gamma_{\infty}=\eta$). This critical value of $\eta$ is obtained as
\begin{eqnarray}
    \eta_{\rm heavy}\sim140\left(\frac{Z}{A}\right)^{1/4}r_{0,5}^{1/4}\,\Gamma_{0}^{3/4}\,\Theta_0,
\end{eqnarray}
by simply setting $r_{\rm M}=r_{\rm ph}$, where the Thomson optical depth is approximated as $\tau_{\rm T}\approx Zn^{\prime}\sigma_{\rm T}r/\Gamma$, taking into account baryon-associated electrons. An outflow with $\eta\gtrsim\eta_{\rm c}$ becomes optically thin before reaching coasting radius (i.e., $r_{\rm M}<r_{\rm ph}$), the coasting Lorentz factor becomes $\eta_{\rm c}$ at $r>r_{\rm M}=\eta_{\rm c}r_{0}$. Therefore, the bulk Lorentz factor evloves as
\beqn
\label{eq:n_e_baryon}
    \Gamma(r)=\Gamma_0\begin{cases}
   r/r_0 & (r<r_{\rm M}) \\
    {\rm min}[\eta,\eta_{\rm heavy}] & (r>r_{\rm M}),
  \end{cases}
\eeqn
where $r_{\rm M}=r_0\,
{\rm min}[\eta,\eta_{\rm heavy}]$.
We consider here the case of relatively high-load FB with $\eta\lesssim10^4$, for which the number density of positrons becomes negligible compared to that of both electrons and baryons (i.e., $n_e^{\prime}\sim Zn^{\prime}$ assuming the charge neutrality). In this case, pair annihilation does not occur anymore and the electron number density conserves:
\beqn
\label{eq:Boltzmann eq}
\partial_r\left(r^2n_e^{\prime}\Gamma \beta\right)=0,
\eeqn
where LHS represents the net pair creation rate.
Therefore, setting $(U^{\prime}+P^{\prime})|_{r=r_0}\sim aT_0^4$ in Eq. \eqref{eq:eta}, the radial evolution of the electron number density may be estimated as
\beqn
\label{eq:n_e_baryon}
    &&n_e^{\prime}(r)\approx \frac{aT_0^4\Gamma_0}{m_pc^2} \left(\frac{Z}{A}\right)\eta^{-1}\nonumber\\
    &&\times\begin{cases}
    \left(r/r_0\right)^{-3} & (r<r_{\rm M}) \\
    {\rm min}[\eta,\eta_{\rm heavy}]^{-1}\left(r/r_0\right)^{-2} & (r>r_{\rm M}).
  \end{cases}
\eeqn
The above evolution is true up to the second critical point with $\eta=\eta_{\rm mild}\sim3.8\times10^4\, (Z/A)\Theta_0\, r_{0,5}$, when $n_e^{\prime}\sim Zn^{\prime}$ at $r=r_{\rm ph}$. The plasma cutoff radius $r_{\rm cutoff}$ at which $\nu_{\rm p}=\nu_{\rm obs}$ is
\beqn
\label{eq:nu_pl_magnetic}
&&r_{\rm cutoff}\sim7.1\times10^{13}\ r_{0,5}\,\Theta_0^2\, \zeta\, \nu_{\rm obs,9}^{-1} \ \ {\rm cm}\nonumber
\\
&&\times\begin{cases} 1 & (\eta<\eta_{\rm heavy}) \\
(\eta/\eta_{\rm heavy})^{-1/2} & (\eta_{\rm heavy}<\eta<\eta_{\rm mild}).
\end{cases}
\eeqn
where we assume $\Gamma_0=1$ and $Z/A\sim1$. 

Although not covered in this work, for completeness, we briefly describe the weak load case. The weakly-loaded baryonic outflow evolution ($\eta>\eta_{\rm mild}$) can be characterized by the additional critical value of $\eta=\eta_{\rm weak}$ when $m_en_e^{\prime}\sim m_pn^{\prime}$ at $r=r_{\rm ph}$ (hence $\eta_{\rm weak}/\eta_{\rm mild}\sim m_p/m_e$). At $\eta_{\rm mild}<\eta$, the effective electron mass can be approximated as ${\tilde{m}}_e\approx(A/2Z)m_e\,{\rm min}\{\eta_{\rm weak}/\eta,1\}$ \citep{nakar05}. 
By replacing $m_e$ with ${\tilde{m}}_e$ in the coasting radius of leptonic outflow $\Gamma_{\infty}\propto m_e^{-1/4}$, the coasting Lorentz factor $\Gamma_{\infty}$ is found to reduce at most by a factor of $(Am_p/2Zm_e)^{1/4}\sim6\,(A/Z)^{1/4}$ compared to the pure leptonic case. The inequalty between $e^{\pm}$ number density does not significantly change the characteristic radii (e.g., $r_{\rm ph}$) that determine the evolution of a quasi-leptonic outflow throughout $\eta>\eta_{\rm heavy}$ \citep{grimsrud98}.

\subsection{Magneto-Leptonic Wind}
If the central engine carries a strong magnetic field, it may significantly contribute to the energy of the relativistic outflow. We consider a cold  magneto-leptonic FB ($P^{\prime}=0$, $U^{\prime}=\rho^{\prime}c^2=n_e^{\prime}m_ec^2$), corresponding to a relativistic limit with high initial magnetization $\sigma_0\gg1$, which is defined by the ratio of Poynting flux to matter energy flux at the magnetosonic point. 
The total energy and mass flux are linked by
\beqn
L=(1+\sigma)\Gamma \dot{M}c^2,
\eeqn
where $(1+\sigma)\Gamma$ is a conserved quantity. In Poynting-flux dominated flows, dissipation of magnetic energy can take place
via a reconnection process. For non-ideal MHD, the dynamical evolution of outflow in relativistic limit is given by \citep{drenkhahn02,ds02}
\beqn
\label{eq:magnetic wind}
\partial_r \Gamma = \frac{2}{c\tau_{\rm dis}}\left(\sigma_0^{3/2}+\sigma_0^{1/2}-\Gamma\right),
\eeqn
where $\tau_{\rm dis}$ is the timescale for dissipation of toroidal magnetic fields. We assume that the complete field decays into kinetic energy.  The timescale for acceleration is solely determined by specific reconnection processes. Here we consider an outflow with stripes of a toroidal magnetic field of alternating polarity (e.g., \citealt{kc84,lyubarsky01}). In this case, the dissipation occurs in the outflow outside the light cylinder with lab-frame timescale
\beqn
\tau_{\rm dis}=\frac{P_{\rm spin}}{\epsilon}\frac{\Gamma^2}{\sqrt{1-\Gamma/\sigma_0^{3/2}}},
\eeqn
where $P_{\rm spin}=3.24$ s is the spin rate of SGR 935+2154 and $\epsilon$ is defined as a fraction of advection velocity of magnetic field lines toward reconnection center with respect to Alfv\'en velocity. \citet{drenkhahn02} showed that the Poynting-flux dominated relativistic flow accelerates as $\Gamma\propto r^{1/3}$ up to the coasting value of $\Gamma_{\rm \infty}=\sigma_0^{3/2}+\sigma_0^{1/2}$ ($\partial_r \Gamma=0$ in Eq. \eqref{eq:magnetic wind}]), which is independent of the reconnection rate $\epsilon$. 
The largest uncertainty lies in the reconnection rate parameter $\epsilon$ and we take $\epsilon=0.1$ as a fiducial value  \citep{drenkhahn02}. 
Simulation studies of reconnecting current sheets suggest a smaller value $\epsilon=0.01$ (e.g., \citealt{uzdensky10}), which may increase the injection radius by about ten times. Nevertheless, due to the relatively slow acceleration $\Gamma\propto r^{1/3}$, this barely affects our final conclusions.

In the absence of dissipation, the bulk Lorentz factor of a magnetised outflow grows as $\Gamma\approx r/r_{\rm lc}$ due to the balance between the electromagnetic and centrifugal forces up to the fast magnetosonic surface, beyond which there is little acceleration \citep{beskin98,contopoulos02,komissarov09}. We set the initial flow velocity to the Alfv\'en four-velocity $u_{\rm A}\equiv B_0^{\prime}/(4\pi\rho_0^{\prime}c^2)^{1/2}=\sigma_0^{1/2}$ (where $B_0^{\prime}$ is the magnetic field and $\rho_0^{\prime}$ is the rest mass
density) at the initial radius $r=r_{\rm lc}\sim10^{10}$ cm. 
Since the dissipation only sets in at $r_{\rm inj}\sim r_{\rm lc}/\epsilon=10\epsilon_{-1}^{-1}\,r_{\rm lc}$, we can safely neglect the dynamical evolution before passing the fast magnetosonic point \citep{drenkhahn02}, unless an extremely high magnetization ($\sigma_0\gg1000$) is considered. 
The initial pair number density is determined at $r=r_{\rm lc}$ by the following condition:
\beqn
E_{\rm flare}\sim(1+\sigma_0)\Gamma_0^2 4\pi\rho_{0}^{\prime}c^2 r_{\rm 0}\,r_{\rm lc}^2,
\eeqn
where $\Gamma_0\approx u_A=\sigma_0^{1/2}$. 
For a cold magnetised outflow, the pair annihilation is negligible and thus the evolution of pair number density is estimated by Eq. \eqref{eq:Boltzmann eq}. For initial magnetizations of $\sigma_0=10$--$1000$, we numerically evaluate the dynamical evolution with Eq. \eqref{eq:magnetic wind} and obtain $r_{\rm cutoff}\sim10^{13}$--$10^{14}$ cm.
For analytic estimate, we use
\beqn
\label{eq:nu_pl_magnetic}
\Gamma\sim\begin{cases}\Gamma_0 & (r_{\rm lc}<r<r_{\rm inj}) \\
\Gamma_0(r/r_{\rm inj})^{1/3} & (r_{\rm inj}<r<r_{\rm sat})
\\
\Gamma_{\infty} & (r_{\rm sat}<r),
    \end{cases}
\eeqn
where $r_{\rm sat}\sim r_{\rm inj}\Gamma_{\infty}^2$ \citep{drenkhahn02} is the saturation radius where the acceleration ends. We confirm that this gives a very good approximation of Lorentz factor during the acceleration phase for $\sigma_0\gg1$. Assuming that the flow is in the acceleration phase, we obtain the cutoff radius for maser-type emission as
\beqn
r_{\rm cutoff}\sim2.9\times10^{14}\ r_{0,5}^{-1/2}\,E_{\rm flare,39}^{1/2}\,  \zeta\,\nu_{\rm obs,9}^{-1}\ \,{\rm cm},
\eeqn
which is remarkably independent of $\sigma_0$ and $\epsilon$.

\bsp	
\label{lastpage}
\end{document}